\documentclass[preprint,12pt]{elsarticle}

\usepackage{amssymb, mathdots}

\journal{XXX}

\begin{document}

\begin{frontmatter}



\title{Construction and enumeration of left dihedral codes satisfying certain duality properties}


\author{Yuan Cao$^{a, \ b}$, Yonglin Cao$^{a, \ \ast}$, Fanghui Ma$^{a}$}

\address{$^{a}$School of Mathematics and Statistics,
Shandong University of Technology, Zibo, Shandong 255091, China
\vskip 1mm $^{b}$Hubei Key Laboratory of Applied Mathematics, Faculty of Mathematics and Statistics, Hubei University, Wuhan 430062, China}
 \cortext[cor1]{corresponding author. \\
 E-mail addresses: ylcao@sdut.edu.cn (Yonglin Cao), \ yuancao@sdut.edu.cn (Yuan Cao), \ fhma@sdut.edu.cn (F. Ma).}

\begin{abstract}
Let $\mathbb{F}_{q}$ be the finite field of $q$ elements and let
$D_{2n}=\langle x,y\mid x^n=1, y^2=1, yxy=x^{n-1}\rangle$ be the dihedral group of order $n$. Left ideals of the group algebra $\mathbb{F}_{q}[D_{2n}]$ are known as
left dihedral codes over $\mathbb{F}_{q}$ of length $2n$, and abbreviated as left $D_{2n}$-codes.
Let ${\rm gcd}(n,q)=1$. In this paper, we give an explicit representation for the Euclidean hull of every
left $D_{2n}$-code over $\mathbb{F}_{q}$. On this basis, we determine all distinct
Euclidean LCD codes and Euclidean self-orthogonal codes which are left $D_{2n}$-codes over $\mathbb{F}_{q}$. In particular, we provide an explicit representation and a precise enumeration for these two subclasses of left $D_{2n}$-codes and self-dual left $D_{2n}$-codes,
 respectively. Moreover, we give a direct and simple method for determining the encoder (generator matrix) of any left $D_{2n}$-code over $\mathbb{F}_{q}$, and present several numerical examples to illustrative our applications.
\end{abstract}


\end{frontmatter}


\section{Introduction}
\label{intro} \noindent
Algebraically structured codes with self-duality and complementary duality are important
families of linear codes that have been extensively studied for both theoretical and practical
reasons (see \cite{JitmaFFA2019}, \cite{Dougherty18,Dougherty19,DoughertyFFA19,Dougherty20}, \cite{Jia2011}, \cite{Nebe2006},
\cite{Saleh2018}, \cite{Willems2002},
\cite{Yang1994}, and references therein).
One-sided ideals of finite group algebras and group rings are called group codes (cf. \cite{Dutra2009}, \cite{Gabriela2015},
\cite{JMC2017}, \cite{Polcino2019},
\cite{Willems2002}).
Since group codes have good parameters (see \cite{Assuena2019}, \cite{Bazzi2006}, \cite{BorelloFFA}, \cite{CaoIEEE2020}, \cite{Dougherty18,Dougherty19,DoughertyFFA19,Dougherty20}, \cite{Fan20}, \cite{Gabriela2015}, \cite{s11}, and references therein), they have been one of the
important sources of constructing good linear codes.

 On the other hand, it could be one of meaningful attempts
to develop resistant post-quantum codes based on error-correcting codes by use of group codes (see \cite{Deundyak2015},
\cite{Deundyak2016}, \cite{Deundyak2018}). Hence it is interesting to study the fine structure of non-abelian group codes
for constructing new classes of codes and developing efficient decoding algorithms.

\par
Since the structure of dihedral groups is the simplest among non-abelian groups, the systematic study of dihedral codes is more important and worthy.
Here, we give a brief survey on known results of dihedral codes as follows:

\par
  $\diamondsuit$ In 2006, Bazzi and Mitter \cite{Bazzi2006} shown that for infinitely
many block lengths a random left ideal in the binary group algebra of
the dihedral group is an asymptotically good rate-half code with
a high probability.

\par
 In 2020, following the ideas of \cite{Bazzi2006}, Borello and Willems \cite{BorelloFFA} proved that group codes over finite fields of any characteristic are asymptotically good. In particular, Fan and Lin \cite{Fan20} constructed asymptotically good Euclidean self-dual dihedral group codes when
the characteristic of the field is even; and constructed both the asymptotically good Euclidean self-orthogonal dihedral
group codes and the asymptotically good Euclidean LCD dihedral group codes when the characteristic of the filed is
odd. \textsf{But these papers did not consider how to determine the specific structure of every dihedral code}.

\par
    $\diamondsuit$ In 2012, McLoughlin \cite{s11}
provided a new construction of the self-dual, doubly-even and extremal [48,24,12]
binary linear block code using a
zero divisor in the group ring $\mathbb{F}_2[D_{48}]$. In recent years, Dougherty et al.
constructed certain good self-dual and formally self-dual codes in \cite{Dougherty18,Dougherty19,DoughertyFFA19,Dougherty20},
using some one-sided ideals of dihedral group rings and other group rings.

\par
  $\diamondsuit$ In 2015, Brochero Mart\'{i}nez \cite{Martinez2015} shown explicitly all central irreducible idempotents and their Wedderburn decomposition of the
dihedral group algebra $\mathbb{F}_q[D_{2n}]$, in the case when every divisor
of $n$ divides $q-1$. In this direction, there are some new results in 2020 and 2021:
\begin{description}
\item{$\triangleright$}
  Vedenev and Deundyak \cite{Vedenev20} studied left ideals in dihedral group algebra by one generalization of the Wedderburn decomposition of $\mathbb{F}_q[D_{2n}]$, where
${\rm gcd}(q,n)=1$, established some connections with the theory of cyclic codes and
obtained some results about code parameters.

\item{$\triangleright$}
  Gao et al. \cite{GaoLetters} given a descriptions to some self-dual group codes with parameters
in the group algebra $\mathbb{F}_q[D_{2n}]$, where $n$ is odd and $q$ is even.

\item{$\triangleright$}
  Gao et al. \cite{GaoDCC} extended the results of \cite{Martinez2015} to the generalized dihedral group algebra
$\mathbb{F}_q[D_{2n,r}]$, where ${\rm gcd}(q,n)=1$ and $q$ is odd, and provided an explicit expression for
primitive idempotents of the group algebra. In particular, some Euclidean LCD codes and Euclidean self-orthogonal
codes in $\mathbb{F}_q[D_{2n,r}]$ were described and counted.

\item{$\triangleright$}
  When ${\rm gcd}(4n, q)= 1$, Gao and Yue \cite{DM21} determined central (right) irreducible idempotents of $\mathbb{F}_q[Q_{4n}]$
in two cases: $q\equiv 1$ (mod $4$) and $q\equiv 3$ (mod $4$), and obtained descriptions and enumerations of Euclidean LCD and self-orthogonal codes in the group algebras $\mathbb{F}_q[Q_{4n}]$.

\item{$\triangleright$}
 Borello and Jamous \cite{Borello20} studied a subclass of dihedral codes. Specifically,
they proved a BCH bound for principal dihedral codes and proposed a definition of principal BCH-dihedral codes.
\end{description}

\noindent
  The above papers are based on radical extensions of the field $\mathbb{F}_q$, the Wedderburn decomposition theory of general semisimple finite group algebra or used the Morita
correspondence (see \cite{Borello20}).
However, these descriptions to dihedral codes are not detailed enough, so that the precise algebraic structure of the dual code for
each dihedral code is no easy to described completely.

 \textsf{In fact, the above descriptions are inconvenient to determine the complete representation and exact enumeration of dihedral codes which satisfy certain duality properties}.

\par
  $\diamondsuit$ The concatenation structure of a linear code is helpful to encode and decode the code efficiently (cf. \cite{s13}).
Hence one way to solve the above problems is to give the concatenation structure of every dihedral code.

   In this direction, some works have been done:
\begin{description}
\item{$\triangleright$}
 In 2016, Cao et al. \cite{CaoFFA2016} provided explicit concatenated structures and precise enumerations to all left dihedral codes, their Euclidean dual codes and the Euclidean self-dual codes in the group algebra $\mathbb{F}_q[D_{2n}]$, for any positive integer $n$ satisfying ${\rm gcd}(q,n)=1$.
  When $q$ is odd and $n$ is even satisfying ${\rm gcd}(n, q)= 1$, Cao et al. \cite{Cao2016} gave concatenated structures and  enumerations for all left quaternion codes and their Euclidean dual codes that are left ideals
of the group algebras $\mathbb{F}_q[Q_{2n}]$, were $Q_{2n}$ is the
generalized quaternion group with $2n$ elements.

\item{$\triangleright$}
 In 2018, Cao et al. \cite{DM2018} extended the results of \cite{CaoFFA2016} to
left dihedral codes over Galois ring ${\rm GR}(p^2,m)$, for any prime $p$ and positive integer $m$.

\item{$\triangleright$}
 In 2020, Cao et al. \cite{CaoIEEE2020} provided an efficiency method to obtain the concatenated structures and precise enumerations to all binary left dihedral codes, their Euclidean dual codes and the self-dual codes in group algebra $\mathbb{F}_2[D_{8m}]$ for any positive integer $m$. In particular, some optimal self-dual binary $[8m,4m]$-codes
 are rediscovered.
\end{description}

   The hull of a linear code plays an important role in determining the complexity of algorithms
for checking permutation equivalence of two linear codes, computing the automorphism group of the code
and in the construction of entanglement-assisted quantum error correcting codes (cf. \cite{Guenda2018}).

  \textsf{As far as we know, the Euclidean hull of any left dihedral code,
the representation and enumeration to all Euclidean LCD left dihedral codes and self-orthogonal left dihedral codes have not been fully studied}.

\par
   The present paper is organized as follows:
   In section 2, We introduce the necessary notation and terminology.
In section 3, we reformulate the results for the concatenated structures of left dihedral codes and their Euclidean dual codes in \cite{CaoFFA2016}, where ${\rm gcd}(n,q)=1$. Then we determine the Euclidean hull of every left $D_{2n}$-code over $\mathbb{F}_q$. In section 4,
we give an explicit representation for all distinct Euclidean LCD codes, Euclidean self-orthogonal codes and Euclidean self-dual codes which are left $D_{2n}$-codes over $\mathbb{F}_{q}$ respectively, and count the precise number of codes in each of these three subclass of left $D_{2n}$-codes.
 In section 5, we give a direct and simple method for determining the encoder (generator matrix) of any left $D_{2n}$-code over $\mathbb{F}_{q}$, and provide several illustrative examples for obtaining self-orthogonal dihedral codes over $\mathbb{F}_{2}$
and LCD dihedral codes over $\mathbb{F}_{3}$ respectively. Section 6 concludes the paper.

\section{Preliminaries}

\noindent
   In this section, we review some necessary concepts and introduce some necessary notation,
which will be used in the following sections.

\par
  Let $\mathbb{F}_{q}$ be the finite field of $q$ elements and $n$ be a positive integer, where $q$ is a power of a prime number. Let
$$D_{2n}=\langle x,y\mid x^n=1, y^2=1, yxy=x^{n-1}\rangle=\{1,x,\ldots,x^{n-1},y,xy,\ldots,x^{n-1}y\}$$
be the dihedral group of order $n$.
The group algebra $\mathbb{F}_{q}[D_{2n}]$ is a vector space over $\mathbb{F}_{q}$ with basis $D_{2n}$. Addition, multiplication with
scalars $c\in \mathbb{F}_{q}$ and multiplication are defined by: for any $a_g,b_g\in \mathbb{F}_{q}$ where $g\in D_{2n}$,
$$\sum_{g\in D_{2n}}a_g g+\sum_{g\in D_{2n}}b_g g=\sum_{g\in D_{2n}}(a_g+b_g) g, \
c(\sum_{g\in D_{2n}}a_g g)=\sum_{g\in D_{2n}}ca_g g,$$
$$(\sum_{g\in D_{2n}}a_g g)(\sum_{g\in D_{2n}}b_g g)=\sum_{g\in D_{2n}}(\sum_{uv=g}a_ub_v)g.$$
Then $\mathbb{F}_{q}[D_{2n}]$ is an associative and noncommutative $\mathbb{F}_{q}$-algebra with identity $1=1_{\mathbb{F}_{q}}1_{D_{2n}}$ where
$1_{\mathbb{F}_{q}}$ and $1_{D_{2n}}$ is the identity elements of $\mathbb{F}_{q}$ and $D_{2n}$ respectively. Readers are referred to \cite{Passman1977}
for more details on group algebra.

\par
   For any $\alpha=(a_{0,0}, a_{1,0},\ldots, a_{n-1,0},a_{0,1}, a_{1,1},\ldots, a_{n-1,1})\in \mathbb{F}_{q}^{2n}$, set
$$
\Psi(\alpha)=a_{0,0}+a_{1,0}x+\ldots+a_{n-1,0}x^{n-1}+a_{0,1}y+a_{1,1}xy+\ldots+a_{n-1,1}x^{n-1}y
$$
Then $\Psi$ is an $\mathbb{F}_{q}$-linear isomorphism from $\mathbb{F}_{q}^{2n}$ onto
$\mathbb{F}_{q}[D_{2n}]$. As in \cite{Gabriela2015}, a nonempty subset $C$ of $\mathbb{F}_q^{2n}$
is called a \textit{left dihedral code} (or \textit{left $D_{2n}$-code} for more precisely) over $\mathbb{F}_q$ if $\Psi(C)$ is a left ideal of the
$\mathbb{F}_q$-algebra $\mathbb{F}_q[D_{2n}]$. From now on, we will identify
$C$ with $\Psi(C)$ and assume ${\rm gcd}(q,n)=1$.

\par
  For any $\alpha=(a_0,a_1,\ldots,a_{2n-1}), \beta=(b_0,b_1,\ldots,b_{2n-1})\in \mathbb{F}_q^{2n}$, the
\textit{Euclidean inner product} of $\alpha$ and $\beta$ is defined by
$[\alpha,\beta]_E=\sum_{i=0}^{2n-1}a_ib_i\in \mathbb{F}_q$.

  Let $\mathcal{C}$ be a linear code of length $2n$ over $\mathbb{F}_q$. The \textit{Euclidean dual code} $\mathcal{C}^{\bot_E}$ and the
\textit{Euclidean hull} ${\rm Hull}_E(\mathcal{C})$ of $\mathcal{C}$ are
defined by
$$
\mathcal{C}^{\bot_E}=\{\alpha\in \mathbb{F}_q^{2n}\mid [\alpha,\beta]_E=0, \ \forall \beta\in \mathcal{C}\}
\ \mbox{and} \ {\rm Hull}_E(\mathcal{C})=\mathcal{C}\cap \mathcal{C}^{\bot_E},
$$
respectively. Moreover, $\mathcal{C}$ is said to be \textit{Euclidean self-dual} (resp. \textit{Euclidean self-orthogonal},
\textit{Euclidean LCD}) if $\mathcal{C}=\mathcal{C}^{\bot_E}$ (resp. $\mathcal{C}\subseteq\mathcal{C}^{\bot_E}$,
$\mathcal{C}\cap\mathcal{C}^{\bot_E}=\{0\}$ and $\mathbb{F}_q^{2n}=\mathcal{C}+\mathcal{C}^{\bot_E}$).

\par
   Let $f(x)$ be a polynomial in $\mathbb{F}_{q}[x]$ satisfying
$f(0)\neq 0$. In this paper, we set
$\widetilde{f}(x)=\widetilde{f(x)}=x^{{\rm deg}(f(x))} f(\frac{1}{x}).$
As in \cite{JitmaFFA2019}, the \textit{reciprocal polynomial}
of $f(x)$ is defined by
$f^{\ast}(x)=(f(x))^{\ast}=f(0)^{-1}\widetilde{f}(x).$
Then $f(x)$ is said to be \textit{self-reciprocal} if $f^{\ast}(x)=f(x)$.

\vskip 3mm\noindent
  {\bf Lemma 2.1} \textit{Let $f(x)$ be a monic irreducible factor of $x^n-1$ in $\mathbb{F}_q[x]$.
Then we have the following conclusions}:

\vskip 2mm\par
  (i) \textit{Let $f(x)$ is self-reciprocal and ${\rm deg}(f(x))=1$}.
\textit{Then $f(x)=x-1$, if $n$ is odd};
\textit{and $f(x)=x\pm 1$, if $n$ is even}.

\vskip 2mm\par
  (ii) \textit{If $f(x)$ is self-reciprocal and ${\rm deg}(f(x))>1$, ${\rm deg}(f(x))$ must be even}.

\vskip 3mm
\par
   The conclusions in Lemma 2.1 can be found in many literatures (see \cite{CaoFFA2016}, for example). In the rest of this paper, we assume
$$x^n-1=f_0(x)f_1(x)\ldots f_r(x)f_{r+1}(x)\ldots f_{r+2t}(x),$$
where $f_0(x)=x-1,f_1(x),\ldots,f_{r+t}(x)$ are pairwise coprime monic irreducible polynomials in $\mathbb{F}_q[x]$ satisfying
the following conditions (cf. Page 205 of \cite{Martinez2015}):
\begin{description}
\item{$\bullet$}
  $r$ and $t$ are nonnegative integers such that $r+2t=n-1$. Set

  $d_i={\rm deg}(f_i(x))$, for all $i=0,1,\ldots,r+2t$.

\item{$\bullet$}
  $f_i^\ast(x)=f_i(x)$, for all $i=0,1,\ldots,r$. Moreover, we have the following:

 when $n$ is odd, $d_i$ is even for all $i=1,\ldots,r$;

 when $n$ is even,
$f_1(x)=x+1$ and $d_i$ is even for all $i=2,\ldots,r$.

\item{$\bullet$}
   $f_{r+j}^\ast(x)=f_{r+j+t}(x)$ and $f_{r+j+t}^\ast(x)=f_{r+j}(x)$, for all $j=1,\ldots,t$.
\end{description}

\noindent
  Then we can write
$$x^n-1=\prod_{i=0}^rf_i(x)\cdot\prod_{j=1}^tf_{r+j}(x)f_{r+j}^\ast(x).$$

\par
   Let $i$ be any integer satisfying $0\leq i\leq r+2t$. We set $F_i(x)=\frac{x^n-1}{f_i(x)}\in \mathbb{F}_q[x]$. Then
${\rm gcd}(F_i(x),f_i(x))=1$. Using Extended Euclidian Algorithm, we can find polynomials
$u_i(x)$, $v_i(x)\in \mathbb{F}_q[x]$ satisfying
\begin{equation}
\label{eq1}
 u_i(x)F_i(x)+v_i(x)f_i(x)=1.
\end{equation}

\noindent
  In the rest of this paper,
we adopt the following notation:

\begin{description}
\item{$\bullet$}
  Let $\mathcal{A}=\frac{\mathbb{F}_q[x]}{\langle x^n-1\rangle}=\{\sum_{j=0}^{n-1}a_jx^j\mid a_0,a_1,\ldots,a_{n-1}\in \mathbb{F}_q\}$
in which the arithmetic is done modulo $x^n-1$. Then $\mathcal{A}$ is an $\mathbb{F}_q$-algebra and a principal
ideal ring. In particular, $x^{-1}=x^{n-1}$ in $\mathcal{A}$.

\item{$\bullet$}
Using Eq. (\ref{eq1}),
  let $\varepsilon_i(x)\in \mathcal{A}$ be defined by
$$
\varepsilon_i(x)\equiv u_i(x)F_i(x)=1-v_i(x)f_i(x) \ (\mbox{mod} \ x^n-1).
$$
By \cite[Lemma 2.1]{Martinez2015}, we know
that
$\varepsilon_i(x)=-\frac{1}{n}\widetilde{((\widetilde{f}_i(x))^\prime)}\frac{x^n-1}{f_i(x)}.$

Set $\mathcal{A}_i=\mathcal{A}\varepsilon_i(x)$ that is the ideal of $\mathcal{A}$
generated by $\varepsilon_i(x)$.

\item{$\bullet$}
  Let $K_i=\frac{\mathbb{F}_q[x]}{\langle f_i(x)\rangle}=\{\sum_{j=0}^{d_i-1}a_jx^j\mid a_0,a_1,\ldots,a_{d_i-1}\in \mathbb{F}_q\}$
in which the arithmetic is done modulo $f_i(x)$.
Then $K_i$ is an extension
field of $\mathbb{F}_q$ with $q^{d_i}$ elements.

In this paper, we regard $K_i$ as a subset of $\mathcal{A}$. But $K_i$ is not a subfield of the ring $\mathcal{A}$,
because their multiplication operations are different.
\end{description}

\par
  Let $l$ be any positive integer. Recall that a linear code $\mathcal{C}$
over $\mathbb{F}_q$ of length $ln$ is called a \textit{$l$-quasi-cyclic code} if
$$(c_{1,n-1},c_{1,0},c_{1,1},\ldots,c_{1,n-2},\ldots,c_{l,n-1},c_{l,0},c_{l,1},\ldots,c_{l,n-2})\in \mathcal{C},$$
for all $\textbf{c}=(c_{1,0},c_{1,1},\ldots,c_{1,n-2},c_{1,n-1},\ldots,c_{l,0},c_{l,1},\ldots,c_{l,n-2},c_{l,n-1})\in \mathcal{C}$.
Now, we identify each codeword $\textbf{c}$ with
$(c_1(x),\ldots,c_l(x))\in \mathcal{A}^l$, where
$$c_j(x)=c_{j,0}+c_{j,1}x+\ldots+c_{j,n-1}x^{n-1}\in \mathcal{A}, \
\forall j=1,\ldots,l.$$
Then $l$-quasi-cyclic codes over $\mathbb{F}_q$ of length $ln$ are the same as
$\mathcal{A}$-submodules of $\mathcal{A}^l$.
Especially, when $l=1$, we know that
\textit{cyclic codes} over $\mathbb{F}_q$ of length $n$ are the same as ideals
of the ring $\mathcal{A}$.

  The following results can be found in many books and literature (cf. \cite[Threorem 4.3.8]{Huffman03},
\cite[Lemma 3.2]{CaoFFA2015}, \cite[Lemma 2.2]{CaoFFA2016}
and \cite[Equation (7)]{CaoIEEE2020}).

\vskip 3mm\noindent
  {\bf Lemma 2.2} (i) \textit{In the ring $\mathcal{A}$, we have that $\sum_{i=0}^{r+2t}\varepsilon_i(x)=1$, $\varepsilon_i(x)^2=\varepsilon_i(x)$ and $\varepsilon_i(x)\varepsilon_j(x)=0$ for
  all integers $i$ and $j$: $0\leq i,j\leq r+2t$ and $i\neq j$}.

\vskip 2mm
  (ii) \textit{In the ring $\mathcal{A}$, we have the following properties}:

\par
   $\diamond$  \textit{$\varepsilon_i(x^{-1})=\varepsilon_i(x)$, for all $i=0,1,\ldots,r$};

\par
   $\diamond$
   \textit{$\varepsilon_{r+j}(x^{-1})=\varepsilon_{r+j+t}(x)$ and $\varepsilon_{r+j+t}(x^{-1})=\varepsilon_{r+j}(x)$,
for all $j=1,\ldots,t$}.

\vskip 2mm
  (iii) \textit{We have $\mathcal{A}=\bigoplus_{i=0}^{r+2t}\mathcal{A}_i$. Moreover, $\mathcal{A}_i$ is a subring
of $\mathcal{A}$ with multiplicative identity $\varepsilon_i(x)$ satisfying $\mathcal{A}_i\cdot \mathcal{A}_j=\{0\}$, for all
$0\leq i\neq j\leq r+2t$}.

\vskip 2mm
  (iv) \textit{For each integer $i$, $0\leq i\leq r+2t$, the following map}
$$\varphi_i: a(x)\mapsto \varepsilon_i(x)a(x) \ ({\rm mod} \ x^n-1), \forall a(x)\in K_i$$
\textit{is an isomorphism of fields from $K_i$ onto $\mathcal{A}_i$. In particular,
we have that $\mathcal{A}_i=\varepsilon_i(x)\cdot K_i$. Here, we think of $K_i$ as a subset of $\mathcal{A}$}.

\vskip 2mm
  (v) \textit{Let $0\leq i\leq r+2t$. Then $\mathcal{A}_i$ is a minimal
cyclic code over $\mathbb{F}_{q}$ of length $n$ with parity check polynomial
$f_i(x)$ and generating idempotent $\varepsilon_i(x)$}.

\vskip 3mm\par
  Now, let $0\leq i\leq r+2t$ and $C_i$ be any linear code over $K_i$ of length $2$, i.e., let $C_i$ be a $K_i$-subspace
of $K_i^2=\{(a(x),b(x))\mid a(x),b(x)\in K_i\}$. For any codeword
$(a(x),b(x))\in C_i$, by Lemma 2.2 (iv) and (v), we know that both $\varphi_i(a(x))$ and $\varphi_i(b(x))$
are codewords in the cyclic code $\mathcal{A}_i$.
Now, define
the \textit{concatenated code} of the inner code $\mathcal{A}_i$ and
the outer code $C_i$ by
$$\mathcal{A}_i\Box_{\varphi_i}C_i = \{(\varphi_i(a(x)),\varphi_i(b(x)))\mid (a(x),b(x))\in C_i\}\subseteq \mathcal{A}_i^2\subset \mathcal{A}^2,$$
where for any $(a(x),b(x))\in C_i$, we have
$$(\varphi_i(a(x)),\varphi_i(b(x)))=\varepsilon_i(x)\cdot (a(x),b(x))=(\varepsilon_i(x)a(x),\varepsilon_i(x)b(x))
\  (\mbox{mod} \ x^n-1).$$

Then we know the following conclusions:
\begin{description}
\item{$\diamond$}
$\mathcal{A}_i\Box_{\varphi_i}C_i$ is a $2$-quasi-cyclic code over $\mathbb{F}_q$ of length $2n$.

\item{$\diamond$}
${\rm dim}_{\mathbb{F}_q}(\mathcal{A}_i\Box_{\varphi_i}C_i)={\rm dim}_{\mathbb{F}_q}(\mathcal{A}_i)\cdot
{\rm dim}_{K_i}(C_i)=d_i\cdot {\rm dim}_{K_i}(C_i)$,\\
 where ${\rm dim}_{K_i}(C_i)$ is the
dimension of $C_i$ over $K_i$, and
$|\mathcal{A}_i\Box_{\varphi_i}C_i|=|C_i|$.

\item{$\diamond$}
$d_{\mathcal{A}_i\Box_{\varphi_i}C_i}\geq d_{\mathcal{A}_i}d_{C_i},$
where

 $d_{\mathcal{A}_i\Box_{\varphi_i}C_i}$ is the minimal Hamming distance of $\mathcal{A}_i\Box_{\varphi_i}C_i$ over $\mathbb{F}_q$,

 $d_{\mathcal{A}_i}$ is minimal Hamming distance of the cyclic code $\mathcal{A}_i$ over $\mathbb{F}_q$,

 $d_{C_i}$ is the minimal Hamming distance of the linear code $C_i$ over $K_i$.
\end{description}

\section{The Euclidean hull of any left $D_{2n}$-code over $\mathbb{F}_{q}$}
\noindent
  In this section, we determine the Euclidean hull ${\rm Hull}_E(\mathcal{C})=\mathcal{C}\cap \mathcal{C}^{\bot_E}$ of any left $D_{2n}$-code $\mathcal{C}$ over $\mathbb{F}_{q}$.
To do this, we need to give a more direct and simpler representation for all distinct left $D_{2n}$-codes and their Euclidean dual codes.

\par
   Since $D_{2n}=\langle x,y\mid x^n=1,y^2=1, yxy=x^{-1}\rangle$, $C^{n}=\langle x\mid x^n=1\rangle$
is a cyclic subgroup of $D_{2n}$ with order $n$ generated by $x$. Obviously, the group algebra $\mathbb{F}_q[C^{n}]$
is the same as the ring $\mathcal{A}=\frac{\mathbb{F}_q[x]}{\langle x^n-1\rangle}$. Hence
$\mathcal{A}$ is the subring of the group algebra $\mathbb{F}_q[D_{2n}]$ and
$$\mathbb{F}_q[D_{2n}]=\{a(x)+b(x)y\mid a(x),b(x)\in \mathcal{A}\}$$
in which $y^2=1$ and $ya(x)=a(x^{-1})y$ for all $a(x)\in \mathcal{A}$, where
$a(x^{-1})=a(x^{n-1})$ (mod $x^n-1$). From this, we deduce
that $\mathbb{F}_q[D_{2n}]$ is a free left $\mathcal{A}$-module with basis $\{1,y\}$. Hence the
map $\Theta$, defined by
$$\Theta: \mathcal{A}^2\rightarrow \mathbb{F}_q[D_{2n}]
\ \mbox{via} \ \Theta: (a(x),b(x))\mapsto a(x)+b(x)y \ (\forall a(x),b(x)\in \mathcal{A}),$$
is an isomorphism of left $\mathcal{A}$-modules from $\mathcal{A}^2$ onto $\mathbb{F}_q[D_{2n}]$.

\par
   Let $\mathcal{C}$ be a nonempty of $\mathbb{F}_q[D_{2n}]$. It is clear that
$\mathcal{C}$ is a left ideal of $\mathbb{F}_q[D_{2n}]$ if and only if $\Theta^{-1}(\mathcal{C})$
is an $\mathcal{A}$-submodule of $\mathcal{A}^2$ and $y\xi\in \mathcal{C}$ for all
$\xi=a(x)+b(x)y\in \mathcal{C}$. Then by $y\xi=b(x^{-1})+a(x^{-1})y$, we
have $\Theta^{-1}(y\xi)=(b(x^{-1}),a(x^{-1}))$. Hence
$\mathcal{C}$ is a left $D_{2n}$-code over $\mathbb{F}_q$ if and only there is a unique
$\mathcal{A}$-submodule $\mathcal{C}^\prime$ of $\mathcal{A}^2$ satisfying the following condition:
$$
(b(x^{-1}),a(x^{-1}))\in \mathcal{C}^\prime, \ \forall (a(x),b(x))\in \mathcal{C}^\prime
$$
such that $\Theta(\mathcal{C}^\prime)=\mathcal{C}$. In the rest of the paper, we will identify
$\mathcal{C}$ with $\mathcal{C}^\prime$. Hence left $D_{2n}$-codes
form an interesting subclass of $2$-qusi-cyclic codes of length $2n$  over $\mathbb{F}_q$.

\vskip 3mm \noindent
  {\bf Theorem 3.1} \textit{Using the notation of Section 2, let $\mathcal{I}_0=\{0\}$ when $n$ is odd; and let $\mathcal{I}_0=\{0,1\}$ when $n$ is even. Then all distinct left $D_{2n}$-codes $\mathcal{C}$ over $\mathbb{F}_q$ and their
Euclidean dual codes $\mathcal{C}^{\bot_E}$
are given by}:
\begin{equation}
\label{eq2}
\mathcal{C}=\bigoplus_{i=0}^{r+2t}(\mathcal{A}_i\Box_{\varphi_i}C_i)=\sum_{i=0}^{r+2t}\varepsilon_i(x)\cdot C_i
\end{equation}
\textit{and}
$$\mathcal{C}^{\bot_E}=\bigoplus_{i=0}^{r+2t}(\mathcal{A}_{i}\Box_{\varphi_{i}}U_{i})
=\sum_{i=0}^{r+2t}\varepsilon_i(x)\cdot U_i$$
\textit{respectively, where $C_i$ and $U_i$ are linear codes of length $2$ over
the finite field $K_i$ with generator matrices $G_i$ and $E_{i}$ respectively. In particular, the matrices $G_i$ and $E_{i}$ are given by the following three cases}:

\vskip 2mm\par
  (i) \textit{For
any integer $i\in \mathcal{I}_0$, we have one of the following two subcases}:

\par
  (i-1) \textit{Let $q$ be odd. Then the pairs $(G_i,E_i)$ of matrices are given by the following table}:
\begin{center}
\begin{tabular}{llll|l}\hline
  $N_i$  &  $G_i$  &   $|C_i|$  &   $d_{C_i}$ & $E_i$ \\ \hline
     1             &  $(0,0)$    &     $1$    &   $0$     &  $I_2$ \\
     1             &  $I_2$  & $q^{2}$ &   $1$     &  $(0,0)$ \\
     1             & $(1,1)$ & $q$  &   $2$     & $(-1,1)$ \\
     1             & $(-1,1)$ & $q$  &  $2$     & $(1,1)$ \\
\hline
\end{tabular}
\end{center}
\textit{where}
\begin{description}
\item
 $\triangleright$ \textit{$N_i$ is the number of pairs $(G_i,E_i)$
in the same row and $I_2=\left(\begin{array}{cc} 1 & 0 \cr 0 & 1\end{array}\right)$}.

\item
 $\triangleright$ \textit{$|C_i|$ is the number of codewords in the linear code $C_i$ over $K_i$}.

\item
 $\triangleright$ \textit{$d_{C_i}$ is the minimal Hamming distance of $C_i$ over $K_i$}.
\end{description}
\noindent
In the below, the meaning of these notation is the same as above.

\par
  (i-2) \textit{If $q$ is even, $\mathcal{I}_0=\{0\}$ and the pairs $(G_0,E_0)$ of matrices are given by the following table}:
\begin{center}
\begin{tabular}{llll|l}\hline
  $N_0$  &  $G_0$  &   $|C_0|$  &   $d_{C_0}$ & $E_0$ \\ \hline
     1             &  $(0,0)$    &     $1$    &   $0$     &  $I_2$ \\
     1             &  $I_2$  & $q^{2}$ &   $1$     &  $(0,0)$ \\
     1             & $(1,1)$ & $q$  &   $2$     & $(1,1)$ \\
\hline
\end{tabular}
\end{center}

\vskip 2mm\par
  (ii) \textit{For any integer $i\in\{1,\ldots,r\}\setminus \mathcal{I}_0$, let $\varrho_i(x)$ be a
primitive element of the finite field $K_i$ with $q^{d_i}$ elements. Then $d_i$ is even and the pairs $(G_i,E_i)$ of matrices are given by the following table}:
\begin{center}
\begin{tabular}{llll|l}\hline
  $N_i$  &  $G_i$  &   $|C_i|$  &   $d_{C_i}$ & $E_i$ \\ \hline
     1             &  $(0,0)$    &     $1$    &   $0$     &  $I_2$ \\
     1             &  $I_2$  & $q^{2d_i}$ &   $1$     &  $(0,0)$ \\
     $q^{\frac{d_i}{2}}+1$   & $(\varrho_i(x)^{s(q^{\frac{d_i}{2}}-1)},1)$
     & $q^{d_i}$  &   $2$    & $(-\varrho_i(x)^{s(q^{\frac{d_i}{2}}-1)},1)$ \\
\hline
\end{tabular}
\end{center}
\textit{where $s$ is any integer satisfying $0\leq s\leq q^{\frac{d_i}{2}}$.}

\vskip 2mm\par
  (iii) \textit{Let $i=r+j$ where $1\leq j\leq t$. Then the sequences $(G_i,G_{i+t}, E_{i+t},E_i)$
of matrices are given by the following table}:
\begin{center}
\begin{tabular}{llll|lll}\hline
  $N_i$  &  $G_i$ &   $|C_i|$  &   $d_{C_i}$ & $G_{i+t}$ & $E_{i+t}$ & $E_i$ \\ \hline
     1   &  $(0,0)$   &     $1$      &   $0$     & $(0,0)$ & $I_2$  & $I_2$ \\
     1   &  $I_2$ & $q^{2d_i}$       &   $1$  & $I_2$ & $(0,0)$ & $(0,0)$ \\
     1   & $(1,0)$ & $q^{d_i}$  &   $1$     &  $(0,1)$ &  $(0,1)$ & $(1,0)$ \\
     $q^{d_i}$   & $(g(x), 1)$ & $q^{d_i}$  &   $2$ &  $(1,g(x^{-1}))$  & $(1,-g(x^{-1}))$ & $(-g(x),1)$ \\
\hline
\end{tabular}
\end{center}
\textit{where}
\begin{description}
\item{$\triangleright$}
 \textit{$N_i$ is the number of sequences $(G_i,G_{i+t}, E_{i+t},E_i)$
in the same row};

\item{$\triangleright$}
 \textit{$|C_{i+t}|=|C_i|$ and $d_{C_{i+t}}=d_{C_i}$};

\item{$\triangleright$}
  \textit{$g(x)=\sum_{l=0}^{d_i-1}a_lx^l$ and
  $g(x^{-1})=g(x^{n-1})=a_0+\sum_{l=1}^{d_i-1}a_lx^{n-l}$} (mod $f_{i+t}(x)$),
  \textit{for any arbitrary elements $a_0,a_1,\ldots,a_{d_i-1}\in \mathbb{F}_q$}.
\end{description}

\par
  \textit{In particular, the Euclidean dual code $\mathcal{C}^{\bot_E}$ of every left $D_{2n}$-code $\mathcal{C}$
is also a left $D_{2n}$-code over $\mathbb{F}_q$}.

\vskip 2mm\par
  \textit{Moreover, the number $\mathcal{N}$ of left $D_{2n}$-codes over $\mathbb{F}_q$ is equal to}
$$\mathcal{N}=\left\{\begin{array}{ll}
 4\cdot \prod_{i=1}^r(q^{\frac{d_i}{2}}+3)\cdot \prod_{j=1}^t(q^{d_{r+j}}+3), & {\rm if} \
 q \ {\rm is} \ {\rm odd} \ {\rm and} \ n \ {\rm is} \ {\rm odd}; \cr
 4^2\cdot \prod_{i=2}^r(q^{\frac{d_i}{2}}+3)\cdot \prod_{j=1}^t(q^{d_{r+j}}+3), & {\rm if} \
 q \ {\rm is} \ {\rm odd} \ {\rm and} \ n \ {\rm is} \ {\rm even}; \cr
 3\cdot \prod_{i=1}^r(q^{\frac{d_i}{2}}+3)\cdot \prod_{j=1}^t(q^{d_{r+j}}+3), & {\rm if} \
 q \ {\rm is} \ {\rm even}.\end{array}\right.$$

\par
  The theorem can be deduced from \cite[Theorem 2.4, Lemma 3.1, Theorem 3.2, Theorem 4.2, Corollary 4.3 and Theorem 5.4]{CaoFFA2016}.
However, the derivation of this theorem from the representation given in \cite{CaoFFA2016} still requires some transformation processes.
Here we give a direct proof by a method paralleling to that used in \cite{CaoIEEE2020}, for the convenience of the reader.

\vskip 3mm\noindent
  {\bf Proof.} For any integer $i$, $0\leq i\leq r+2t$,
and $(\xi_{i,0}(x),\xi_{i,1}(x))\in K_i^2$, we define
\begin{eqnarray*}
&&\Phi((\xi_{0,0}(x),\xi_{0,1}(x)),(\xi_{1,0}(x),\xi_{1,1}(x)),\ldots,(\xi_{r+2t,0}(x),\xi_{r+2t,1}(x))) \\
&=&\sum_{i=0}^{r+2t}(\varphi_i(\xi_{i,0}(x)),\varphi_i(\xi_{i,1}(x))) \\
&=&(\sum_{i=0}^{r+2t}\varepsilon_i(x)\cdot\xi_{i,0}(x),\sum_{i=0}^{r+2t}\varepsilon_i(x)\cdot\xi_{i,1}(x)) \
 ({\rm mod} \ x^n-1).
\end{eqnarray*}
Then by Lemma 2.2 (i), (iii) and (iv), we see that the map $\Phi$ is an
isomorphism of $\mathcal{A}$-modules from $K_0^2\times K_1^2\ldots\times K_{r+2t}^2$
onto $\mathcal{A}^2$. Now, let $\mathcal{C}$ be any $\mathcal{A}$-submodule
of $\mathcal{A}^2$. Then for each integer $i$, $0\leq i\leq r+2t$, there is a unique $K_i$-subspace $C_i$ of $K_i^2$ such that
\begin{equation}
\label{eq3}
\mathcal{C}=\Phi(C_0\times C_1\times\ldots\times C_{r+2t})
=\sum_{i=0}^{r+2t}\varepsilon_i(x)\cdot C_i
=\bigoplus_{i=0}^{r+2t}(\mathcal{A}_i\Box_{\varphi_i}C_i).
\end{equation}

\par
  Let $(a(x),b(x))\in \mathcal{C}$. Then we have
$$a(x)=\sum_{i=0}^{r+2t}\varepsilon_i(x)\xi_{i,0}(x) \ \mbox{and} \ b(x)=\sum_{i=0}^{r+2t}\varepsilon_i(x)\xi_{i,1}(x),$$
where $(\xi_{i,0}(x),\xi_{i,1}(x))\in C_i$ for all $i=0,1,\ldots,r+2t$. By Lemma 2.2 (ii), we have that $\varepsilon_i(x^{-1})=\varepsilon_i(x)$ for all $i=0,1,\ldots,r$,
$\varepsilon_{i}(x^{-1})=\varepsilon_{i+t}(x)$ and $\varepsilon_{i+t}(x^{-1})=\varepsilon_{i}(x)$
for all $i=r+1,\ldots,r+t$ in the ring $\mathcal{A}$. Then using Lemma 2.2 (iv), we obtain
$$a(x^{-1})=\sum_{i=0}^{r}\varepsilon_i(x)\xi_{i,0}(x^{-1})
+\sum_{i=r+1}^{r+t}(\varepsilon_{i}(x)\xi_{i+t,0}(x^{-1})+\varepsilon_{i+t}(x)\xi_{i,0}(x^{-1})),$$
$$b(x^{-1})=\sum_{i=0}^{r}\varepsilon_i(x)\xi_{i,1}(x^{-1})
+\sum_{i=r+1}^{r+t}(\varepsilon_{i}(x)\xi_{i+t,1}(x^{-1})+\varepsilon_{i+t}(x)\xi_{i,1}(x^{-1})),$$
From these, by Lemma 2.2 (i) (iii) and Eq. (\ref{eq3}), we deduce that
\begin{eqnarray*}
(b(x^{-1}),a(x^{-1}))\in \mathcal{C}\Leftrightarrow
\left\{\begin{array}{ll}(\xi_{i,1}(x^{-1}),\xi_{i,0}(x^{-1}))\in C_i, & 0\leq i\leq r;\cr
(\xi_{i+t,1}(x^{-1}),\xi_{i+t,0}(x^{-1}))\in C_{i}, & r+1\leq i\leq r+t; \cr
(\xi_{i,1}(x^{-1}),\xi_{i,0}(x^{-1}))\in C_{i+t}, & r+1\leq i\leq r+t.\end{array}\right.
\end{eqnarray*}
Since $C_i$ is a $K_i$-subspace of $K_i^2$, i.e., $C_i$ is a linear code over $K_i$ of length $2$, we must have
that ${\rm dim}_{K_i}(C_i)\in \{0,1,2\}$. Now, let's discuss by two cases: when $0\leq i\leq r$;
when $r+1\leq i\leq r+2t$.

\par
  ($\dag$) Let $0\leq i\leq r$. Obviously, $C_i$ satisfies the above condition if $C_i=\{0\}$
or $C_i=K_i^2$. Further, when $C_i=\{0\}$, $G_i=(0,0)$ is a generator matrix of $C_i$; when $C_i=K_i^2$,
$G_i=I_2$ is a generator matrix of $C_i$.

\par
  Let ${\rm dim}_{K_i}(C_i)=1$. Since $C_i$ satisfies $(\xi_{i,1}(x^{-1}),\xi_{i,0}(x^{-1}))\in C_i$
for all $(\xi_{i,0}(x),\xi_{i,1}(x))\in C_i$, there exists a unique element $g(x)\in K_i^\times$ such
that $G_i=(g(x),1)$ is a generator matrix of the linear code $C_i$. Moreover, we see that $(1,g(x^{-1}))\in C_i$ if and only
if $(1,g(x^{-1}))=u(x)(g(x),1)$ for some $u(x)\in K_i$, and the latter is equivalent to
that $g(x)g(x^{-1})=1$. Then we need to consider the following two subcases.

\par
  ($\dag$-1) Let $i\in \mathcal{I}_0$. Then $K_i=\mathbb{F}_q$. In this case, we have
that $g(x)=c$ for some $c\in \mathbb{F}_q^\times$ and the condition $g(x)g(x^{-1})=1$ is simplified
to $c^2=1$. Hence, we have the following conclusions:

\par
  $\diamond$ If $q$ is odd, $c=\pm 1$. Therefore, there are two codes $C_i$ with generator matrices
$G_i=(1,1)$ and $G_i=(-1,1)$ respectively.

\par
  $\diamond$ If $q$ is even, $c=1$. Therefore, there is only one code $C_i$ with generator matrix
$G_i=(1,1)$.

\par
  ($\dag$-2) Let $i\in \{1,2,\ldots,r\}\setminus\mathcal{I}_0$. By Lemma 2.1 (ii), $f_i(x)$
is an irreducible self-reciprocal polynomial in $\mathbb{F}_q[x]$ of even degree $d_i\geq 2$. Then
we have that $x^{-1}=x^{q^{\frac{d_i}{2}}}$ in the finite field $K_i=\frac{\mathbb{F}_q[x]}{\langle f_i(x)\rangle}$.
This implies $g(x^{-1})=g(x)^{q^{\frac{d_i}{2}}}$. Since
$\varrho_i(x)$ is a primitive element of $K_i$ with multiplicative order $q^{d_i}-1$,
the condition $g(x)g(x^{-1})=1$, i.e., $g(x)^{q^{\frac{d_i}{2}}+1}=1$, is equivalent to that
$$g(x)=\varrho_i(x)^{s(q^{\frac{d_i}{2}}-1)}, \ s=0,1,\ldots,q^{\frac{d_i}{2}}.$$

\par
  ($\ddag$) Let $i=r+j$ where $1\leq j\leq t$. Then the pair $(G_i,G_{i+t})$ of codes satisfies the above conditions if and only if the
codes $G_i$ and $G_{i+t}$ are given as following:

\par
  $\triangleright$ $C_i$ is an arbitrary linear code over $K_i$ of length $2$. Hence a generator matrix $G_i$ of
$C_i$ is one of the following $q^{d_i}+1$ matrices:
$$G_i=(1,0), \ G_i=(g(x),1) \ \mbox{where} \ g(x)\in K_i.$$

\par
  $\triangleright$ Let $C_{i+t}=\{(\xi_{i,1}(x^{-1}),\xi_{i,0}(x^{-1}))\mid (\xi_{i,0}(x),\xi_{i,1}(x))\in C_i\}$.
Hence a generator matrix $G_{i+t}$ of
$C_{i+t}$ is one of the following $q^{d_i}+1$ matrices:
$$G_{i+t}=(0,1) \ \mbox{if} \ G_i=(1,0), \ G_{i+t}=(1,g(x^{-1}))  \ \mbox{if} \ G_i=(g(x),1),$$
where $g(x^{-1})\in K_{i+t}$, i.e., $g(x^{-1})=g(x^{n-1})$ (mod $f_{i+t}(x)$), for
any polynomial $g(x)=\sum_{l=0}^{d_i-1}a_lx^l\in K_i$ with $a_l\in \mathbb{F}_q$.

\par
  As mentioned above, we conclude that all distinct left $D_{2n}$-codes $\mathcal{C}$ over $\mathbb{F}_q$
have been given by this theorem.

\par
  Finally, let $\mathcal{D}=\bigoplus_{i=0}^{r+2t}(\mathcal{A}_{i}\Box_{\varphi_{i}}U_{i})=\sum_{i=0}^{r+2t}\varepsilon_i(x)\cdot U_i$ be given by this theorem. Specifically, $U_i$ is a linear code over $K_i$ of length $2$ with
a generator matrices $E_i$, $0\leq i\leq r+2t$. We set
$\Upsilon(\mathcal{D})=\{\left(\begin{array}{c}u(x^{-1})\cr v(x^{-1})\end{array}\right)\mid (u(x),v(x))\in \mathcal{D}\}$.
Then we have
$\Upsilon(\mathcal{D})=\sum_{i=0}^{r+2t}\varepsilon_i(x^{-1})\cdot \Upsilon(U_i)$,
where
$$\Upsilon(U_i)=\{\left(\begin{array}{c}a_i(x^{-1})\cr b_i(x^{-1})\end{array}\right)\mid (a_i(x),b_i(x))\in U_i\}, \
\forall i: 0\leq i\leq r+2t.$$
By the tables of the theorem, we have that
$G_i\cdot \Upsilon(E_i)=0$ for all $i=0,1,\ldots,r$,
$G_i\cdot \Upsilon(E_{i+t})=0$ and $G_{i+t}\cdot \Upsilon(E_i)=0$ for $i=r+1,\ldots,r+t$.
From these and by Lemma 2.2 (i)--(iv), one can easily verify that $|\mathcal{C}||\mathcal{D}|=\prod_{i=0}^{r+2t}|C_i|\prod_{i=0}^{r+2t}|U_{i}|
=q^{\sum_{i=0}^{r+2t}2d_i}=q^{2n}=|\mathbb{F}_q^{2n}|$ and
\begin{eqnarray*}
\mathcal{C}\cdot \Upsilon(\mathcal{D})
 &=&\sum_{i=0}^{r}\varepsilon_i(x)(C_i\cdot \Upsilon(U_{i}))+\sum_{i=r+1}^{r+t}\left(\varepsilon_i(x)(C_i\cdot \Upsilon(U_{i+t}))
  +\varepsilon_{i+t}(x)(C_{i+t}\cdot \Upsilon(U_{i}))\right)\\
 &\equiv& \{0\} \ ({\rm mod} \ x^n-1)
\end{eqnarray*}

   As stated above, we conclude that $\mathcal{C}^{\bot_E}=\mathcal{D}$.
\hfill $\Box$

\vskip 3mm \par
  From now on, the representation of any left $D_{2n}$-code $\mathcal{C}$ over $\mathbb{F}_q$ given by Theorem 3.1
is called the \textit{canonical representation} of $\mathcal{C}$.
Then we can determine the Euclidean hull of any arbitrary left $D_{2n}$-code.

\vskip 3mm \noindent
  {\bf Theorem 3.2} \textit{Let $\mathcal{C}$ be any left $D_{2n}$-code over $\mathbb{F}_q$ with the canonical representation  $\mathcal{C}=\bigoplus_{i=0}^{r+2t}(\mathcal{A}_i\Box_{\varphi_i}C_i)$
 given by Theorem 3.1. Then
the Euclidean hull ${\rm Hull}_E(\mathcal{C})$ of $\mathcal{C}$
is also a left $D_{2n}$-code over $\mathbb{F}_q$}.

   \textit{In particular, the canonical representation of
${\rm Hull}_E(\mathcal{C})$ is given as below}:
$${\rm Hull}_E(\mathcal{C})=\bigoplus_{i=0}^{r+2t}(\mathcal{A}_i\Box_{\varphi_i}\Omega_i),$$
\textit{where $\Omega_{i}$ is a linear code of length $2$ over
$K_i$ with generator matrix $M_{i}$, $0\leq i\leq r+2t$, and the matrices $M_{i}$ are given by the following three cases}:

\vskip 2mm\par
  (i) \textit{Let $i\in \mathcal{I}_0$. We have one of the following two subcases}:

\par
  (i-1) \textit{If $q$ is odd, the matrix $M_i$ is given by}:
  $$M_i=(0,0), \ \forall G_i\in \left\{(0,0), I_2, (1,1), (-1,1)\right\}.$$
\par
  (i-2) \textit{If $q$ is even, $\mathcal{I}=\{0\}$ and the matrix $M_0$ is given by}:
$$M_0=\left\{\begin{array}{ll} (0,0), & {\rm if} \ G_0\in \{(0,0),I_2\}; \cr
  (1,1), & {\rm if} \ G_0=(1,1). \end{array}\right.$$

\vskip 2mm\par
  (ii) \textit{Let $i\in \{1,\ldots,r\}\setminus \mathcal{I}_0$. The matrix $M_i$ is given by one the following two cases}:

\par
  $\diamondsuit$ \textit{When $q$ is odd, we have}
$$M_i=(0,0), \ \forall G_i\in \{(0,0), I_2\}\cup \left\{(\varrho_i(x)^{s(q^{\frac{d_i}{2}}-1)},1)\mid 0\leq s\leq q^{\frac{d_i}{2}}\right\}.$$

\par
  $\diamondsuit$ \textit{When $q$ is even, we have}
$$M_i=\left\{\begin{array}{ll} (0,0), & {\rm if} \ G_i\in \{(0,0),I_2\}; \cr
  (\varrho_i(x)^{s(q^{\frac{d_i}{2}}-1)},1), & {\rm if} \ G_i=(\varrho_i(x)^{s(q^{\frac{d_i}{2}}-1)},1),
  \ 0\leq s\leq q^{\frac{d_i}{2}}. \end{array}\right.$$

\vskip 2mm\par
  (iii) \textit{Let $i=r+j$ where $1\leq j\leq t$. Then the pairs $(M_i,M_{i+t})$ of matrices
are given by one the following two cases}:

\par
  $\diamondsuit$ \textit{When $q$ is odd, we have}
\begin{description}
\item
 $(M_i,M_{i+t})=\left\{\begin{array}{ll} ((0,0),(0,0)), & {\rm if} \ G_i\in \{(0,0),I_2\}; \cr
  ((1,0), (0,1)),  & {\rm if} \ G_i=(1,0); \cr
  ((0,1), (1,0)),  & {\rm if} \ G_i=(0,1).\end{array}\right.$

\item
  \textit{$M_i=M_{i+t}=(0,0)$, if $G_i=(g(x), 1)$ where $g(x)=\sum_{l=0}^{d_i-1}a_lx^l$ for
any $a_0,a_1,\ldots,a_{d_i-1}\in \mathbb{F}_q$ satisfying $(a_0,a_1,\ldots,a_{d_i-1})\neq (0,\ldots,0)$}.
\end{description}

\par
  $\diamondsuit$ \textit{When $q$ is even, we have}
\begin{description}
\item
  $M_i=M_{i+t}=(0,0)$, if $G_i\in \{(0,0),I_2\}$;

\item
  $M_i=(1,0)$ and $M_{i+t}=(0,1)$, if $G_i=(1,0)$;

\item
  \textit{$M_i=(g(x),1)$ and $M_{i+t}=(1,g(x^{-1}))$, if $G_i=(g(x), 1)$ where $g(x)=\sum_{l=0}^{d_i-1}a_lx^l$
  and $g(x^{-1})=a_0+\sum_{l=1}^{d_i-1}a_lx^{n-l}$ $({\rm mod} \ f_{i+t}(x))$
  for any elements $a_0,a_1,\ldots,a_{d_i-1}\in \mathbb{F}_q$}.
\end{description}

\noindent
  {\bf Proof.} Let $\mathcal{C}^{\bot_E}$ be the Euclidean dual code of $\mathcal{C}$.
By Theorem 3.1, we have
$$\mathcal{C}=\sum_{i=0}^{r+2t}\varepsilon_i(x) C_i
\ {\rm and} \ \mathcal{C}^{\bot_E}=\sum_{i=0}^{r+2t}\varepsilon_{i}(x)U_{i}$$
where $C_i$ and $U_{i}$ are linear codes of length $2$ over
$K_i$ with generator matrices $G_i$ and $E_{i}$ respectively. Then by Lemma 2.2 (i)--(iv), we conclude that
$${\rm Hull}_E(\mathcal{C})=\mathcal{C}\cap \mathcal{C}^{\bot_E}
=\sum_{i=0}^{r+2t}\varepsilon_i(x) \Omega_i
=\bigoplus_{i=0}^{r+2t}\mathcal{A}_i\Box_{\varphi_i}\Omega_i,$$
where $\Omega_i=(C_i\cap U_i)$ which is a linear code of length $2$
over the finite field $K_i$, for all integers $i$: $0\leq i\leq r+2t$. Let $M_i$ be a generator matrix of $\Omega_i$. Since both $C_i$ and $U_i$ are linear code of length $2$
over $K_i$, ${\rm dim}_{K_i}(C_i)+{\rm dim}_{K_i}(U_i)=2$, $G_i$ is a generator matrix of $G_i$, $E_i$ is a generator matrix of $U_i$, the matrices
$G_i$ and $E_i$ are given by Theorem 3.1, we deduce the following conclusions:

 $\diamond$ $M_i=\left\{\begin{array}{ll} G_i, & {\rm if} \ G_i=E_i; \cr
    (0,0), & {\rm otherwise}, \end{array}\right.$ for all $i=0,1,\ldots, r$.

 $\diamond$ $(M_i, M_{i+t})=\left\{\begin{array}{ll} (G_i,G_{i+t}) & {\rm if} \ G_i=E_i; \cr
    ((0,0),(0,0)) & {\rm otherwise}, \end{array}\right.$ for all $i=r+1,\ldots,r+t$.

\noindent
Then the conclusion of this theorem can be derived from Theorem 3.1 immediately.
Here we omit the details. \hfill $\Box$

\section{The LCD left $D_{2n}$-codes and self-orthogonal left $D_{2n}$-codes}
\noindent
In this section, we give an explicit representation and a precise enumeration for all distinct Euclidean LCD codes and Euclidean self-orthogonal codes which are left $D_{2n}$-codes over $\mathbb{F}_{q}$ respectively.
At the end, in order to compare the enumerations of the classes of left $D_{2n}$-codes satisfying certain dual properties, we reformulate the representation of self-dual left $D_{2n}$-codes over $\mathbb{F}_q$ given by \cite[Corollary 5.5]{CaoFFA2016},
when $q$ is even.

\vskip 3mm\par
   In the following, we determine Euclidean LCD left $D_{2n}$-codes over $\mathbb{F}_q$.
Since $\mathbb{F}_q$ is a finite field, we see that
a left $D_{2n}$-code $\mathcal{C}=\bigoplus_{i=0}^{r+2t}(\mathcal{A}_i\Box_{\varphi_i}C_i)$ over $\mathbb{F}_q$ is Euclidean LCD if and only if ${\rm Hull}_E(\mathcal{C})=\{0\}$. Using Theorem 3.2, the latter condition
is equivalent to that
$$M_i=(0,0), \ \forall i=0,1,\ldots,r+2t.$$
Then we consider two situations: when $q$ is even; and when $q$ is odd.

\par
  ($\dag$) When $q$ is even, by Theorem 3.2, all distinct Euclidean LCD left $D_{2n}$-codes over $\mathbb{F}_q$ are given as follows:
$$\mathcal{C}=\bigoplus_{i=0}^{r+2t}(\mathcal{A}_i\Box_{\varphi_i}C_i),$$
where $C_i$ is a linear code over $K_i$ of length $2$ with generator matrix $G_i$,
$0\leq i\leq r+2t$, and $G_i$ is given by the follows:

\par
  $\diamond$ If $0\leq i\leq r$, $G_i\in \{0,I_2\}$.

\par
  $\diamond$ If $i=r+j$ where $1\leq j\leq t$, $G_i=G_{i+t}\in \{0,I_2\}$.

\noindent
  Therefore, the number of Euclidean LCD left $D_{2n}$-codes over $\mathbb{F}_q$ is equal to
$$\mathcal{N}_{{\rm E-LCD}}=2^{1+r+t}.$$

\par
 ($\ddag$) When $q$ is odd, using Theorem 3.2, we arrive at the following:

\vskip 3mm\noindent
  {\bf Theorem 4.1} \textit{Let $q$ be odd. Using the notation of Theorem 3.1, all distinct Euclidean LCD left $D_{2n}$-codes over $\mathbb{F}_q$ are given as follows}:
$$\mathcal{C}=\bigoplus_{i=0}^{r+2t}(\mathcal{A}_i\Box_{\varphi_i}C_i),$$
\textit{where $C_i$ is a linear code over $K_i$ of length $2$ with generator matrix $G_i$,
$i\leq i\leq r+2t$, and $G_i$ is given by the following three cases}:

\vskip 2mm\par
  (i) \textit{When $i\in \mathcal{I}_0$, we have that} $G_i\in \left\{(0,0), I_2, (1,1), (-1,1)\right\}.$

\vskip 2mm\par
  (ii) \textit{When $i\in \{0,1,\ldots,r\}\setminus \mathcal{I}_0$, we have that}
$$G_i\in \{(0,0), I_2\}\cup \left\{(\varrho_i(x)^{s(q^{\frac{d_i}{2}}-1)},1 )\mid 0\leq s\leq q^{\frac{d_i}{2}}\right\}.$$

\vskip 2mm\par
  (iii) \textit{Let $i=r+j$ where $1\leq j\leq t$. Then the pairs $(G_i,G_{i+t})$ of
matrices are given by the following two cases}:

\par
  $\flat$) $G_i=G_{i+t}\in \{(0,0),I_2\}$;

\par
  $\natural$) \textit{$G_i=(g(x), 1)$ and $G_{i+t}=(1,g(x^{-1}))$,
where $g(x)=\sum_{l=0}^{d_i-1}a_lx^l$ and
$g(x^{-1})=a_0+\sum_{l=1}^{d_i-1}a_lx^{n-l} \ ({\rm mod} \ f_{i+t}(x))$
for any $a_0,a_1,\ldots, a_{d_i-1}\in \mathbb{F}_q$ satisfying
$(a_0,a_1,\ldots, a_{d_i-1})\neq (0,0,\ldots,0)$}.

\vskip 2mm\par
  \textit{Therefore, the number $\mathcal{N}_{{\rm E-LCD}}$ of Euclidean LCD left $D_{2n}$-codes over $\mathbb{F}_q$ is equal to}
$$\mathcal{N}_{{\rm E-LCD}}=\left\{\begin{array}{ll}
 4\cdot \prod_{i=1}^r(q^{\frac{d_i}{2}}+3)\cdot \prod_{j=1}^t(q^{d_{r+j}}+1), & {\rm if} \
 \ n \ {\rm is} \ {\rm odd}; \cr
 4^2\cdot \prod_{i=2}^r(q^{\frac{d_i}{2}}+3)\cdot \prod_{j=1}^t(q^{d_{r+j}}+1), & {\rm if} \
 n \ {\rm is} \ {\rm even}.\end{array}\right.$$

\par
  \textit{In particular, every left $D_{2n}$-codes over $\mathbb{F}_q$ is an Euclidean LCD code when each monic
irreducible divisor of $x^n-1$ in $\mathbb{F}_q[x]$ is self-reciprocal}.

\vskip 3mm\noindent
  {\bf Proof.} Let $\mathcal{C}$ be a left $D_{2n}$-code over $\mathbb{F}_q$ with the canonical representation
$\mathcal{C}=\bigoplus_{i=0}^{r+2t}(\mathcal{A}_i\Box_{\varphi_i}C_i)$ given by Theorem 3.1. Then its Euclidean hull is
${\rm Hull}_E(\mathcal{C})=\bigoplus_{i=0}^{r+2t}(\mathcal{A}_i\Box_{\varphi_i}\Omega_i)$, where
$\Omega_i$ is a linear code over $K_i$ of length $2$ with a generator matrix $M_i$ and $M_i$
is determined by Theorem 3.2.

\par
  It is clear that $\mathcal{C}$ is an Euclidean LCD code over $\mathbb{F}_q$ if and only if
${\rm Hull}_E(\mathcal{C})$ $=\{0\}$. The latter is equivalent to that the code $C_i$ satisfies $C_i\cap U_i=\Omega_i=\{0\}$, i.e.,
the generator matrix $G_i$ of $C_i$ satisfies $M_i=(0,0)$, for all integers $i$: $0\leq i\leq r+2t$. Then by Theorems 3.1 and 3.2, one can easily deduce the conclusions
for the representation of Euclidean LCD left $D_{2n}$-codes over $\mathbb{F}_q$.

\par
  Finally, let every monic
irreducible divisor of $x^n-1$ in $\mathbb{F}_q[x]$ be self-reciprocal, i.e., $t=0$.
By Theorem 3.1, we see that the number of LCD left $D_{2n}$-codes over $\mathbb{F}_q$ is the same as
the number of all left $D_{2n}$-codes over $\mathbb{F}_q$.
In this case, every left $D_{2n}$-codes over $\mathbb{F}_q$ is an Euclidean LCD code.
\hfill $\Box$

\vskip 3mm\noindent
  {\bf Remark} Let $q$ be odd. By Theorems 3.1 and 4.1, we have that
$$\frac{\mathcal{N}_{{\rm E-LCD}}}{\mathcal{N}}=\prod_{j=1}^t\left(1-\frac{2}{q^{d_{r+j}}+3}\right)\approx 1,$$
if all $q^{d_{r+j}}$ are large enough when $t\geq 1$.

  Hence a left $D_{2n}$-code over $\mathbb{F}_q$ has a higher probability that is an Euclidean LCD code,
when $q$ is odd (and large enough if $t\geq 1$).

\vskip 3mm\par
  Now, we determine Euclidean self-orthogonal left $D_{2n}$-codes over $\mathbb{F}_q$.

\vskip 3mm \noindent
  {\bf Theorem 4.2} \textit{Using the notation of Theorem 3.1, all distinct Euclidean self-orthogonal left $D_{2n}$-codes over $\mathbb{F}_q$ are given as follows}:
$$\mathcal{C}=\bigoplus_{i=0}^{r+2t}(\mathcal{A}_i\Box_{\varphi_i}C_i),$$
\textit{where $C_i$ is a linear code over $K_i$ of length $2$ with generator matrix $G_i$,
$i\leq i\leq r+2t$, and $G_i$ is given by the following three cases}:

\vskip 2mm\par
  (i) \textit{Let $i\in \mathcal{I}_0$. We have one of the following two subcases}:

\par
  (i-1) \textit{When $q$ is odd, $G_i=(0,0)$}.

\par
  (i-2) \textit{When $q$ is even, $\mathcal{I}_0=\{0\}$ and $G_0\in\{(0,0), (1,1)\}$}.

\vskip 2mm\par
  (ii) \textit{Let $i\in \{1,\ldots,r\}\setminus \mathcal{I}_0$. Then the matrix $G_i$ is given by one the following two cases}:

\par
  $\triangleright$ \textit{When $q$ is odd, $G_i=(0,0)$}.

\par
  $\triangleright$ \textit{When $q$ is even, $G_i\in \{(0,0)\}\cup \{(\varrho_i(x)^{s(q^{\frac{d_i}{2}}-1)},1)\mid
  0\leq s\leq q^{\frac{d_i}{2}}\}$}.

\vskip 2mm\par
  (iii) \textit{Let $i=r+j$ where $1\leq j\leq t$. Then the pairs $(G_i,G_{i+t})$ of matrices
are given by one the following two cases}:

\par
  $\triangleright$ \textit{If $q$ is odd, $(G_i,G_{i+t})\in \{((0,0),(0,0)),((1,0), (0,1)),((0,1), (1,0))\}$}.

\par
  $\triangleright$ \textit{If $q$ is even, the pairs $(G_i,G_{i+t})$ of matrices are give by the following three subcases}:

\par
   $\flat$) $G_i=G_{i+t}=(0,0)$;

\par
   $\natural$) \textit{$G_i=(1,0)$ and $G_{i+t}=(0,1)$};

\par
   $\sharp$) \textit{$G_i=(g(x),1)$ and $G_{i+t}=(1,g(x^{-1}))$, where
$g(x)=\sum_{l=0}^{d_i-1}a_lx^l$ and $g(x^{-1})=a_0+\sum_{l=1}^{d_i-1}a_lx^{n-l} \ ({\rm mod} \ f_{i+t}(x))$
for any $a_0,a_1,\ldots, a_{d_i-1}\in \mathbb{F}_q$}.

\vskip 2mm\par
 \textit{Let $\mathcal{N}_{E-SO}$ be the number of self-orthogonal left $D_{2n}$-codes over $\mathbb{F}_q$. Then}
$$\mathcal{N}_{E-SO}=\left\{\begin{array}{ll}
 3^t, & {\rm if} \
 q \ {\rm is} \ {\rm odd}; \cr
 2\cdot \prod_{i=1}^r(q^{\frac{d_i}{2}}+2)\cdot \prod_{j=1}^t(q^{d_{r+j}}+2), & {\rm if} \
 q \ {\rm is} \ {\rm even}.\end{array}\right.$$

\noindent
  {\bf Proof.} Let $\mathcal{C}=\bigoplus_{i=0}^{r+2t}(\mathcal{A}_i\Box_{\varphi_i}C_i)$ be a left $D_{2n}$-code over $\mathbb{F}_q$ given by Theorem 3.1. Then $\mathcal{C}$ is self-orthogonal if and only if
$\mathcal{C}\subseteq \mathcal{C}^{\bot_E}$, i.e.,
${\rm Hull}_E(\mathcal{C})=\mathcal{C}\cap \mathcal{C}^{\bot_E}=\mathcal{C}$. Using the notation of Theorem 3.2,
we see that the latter condition is equivalent to $G_i=M_i$ for all $i=0,1,\ldots,r+2t$. From this and by Theorems 3.1 and 3.2,
one can easily obtain the conclusions. Here we omit the details.
\hfill $\Box$

\vskip 3mm\noindent
  {\bf Remark} Let $q$ be even. By Theorems 3.1 and 4.2, we have that
$$\frac{\mathcal{N}_{{\rm E-SO}}}{\mathcal{N}}=\frac{2}{3}\cdot \prod_{i=1}^r\left(1-\frac{1}{q^{d_{i}}+3}\right)
\cdot\prod_{j=1}^t\left(1-\frac{1}{q^{d_{r+j}}+3}\right)\approx \frac{2}{3},$$
if all $q^{d_{i}}$ and $q^{d_{r+j}}$ are large enough.

  Hence the probability of a left $D_{2n}$-code over $\mathbb{F}_q$ being Euclidean self-orthogonal is approximately equal to $\frac{2}{3}$,
when $q$ is even (and large enough).

\vskip 3mm \par
  At the end of this section, we reformulate the concatenated structure of Euclidean self-dual left $D_{2n}$-codes over $\mathbb{F}_q$
given by \cite[Corollary 5.5]{CaoFFA2016}. In fact, from the condition: $\mathcal{C}=\mathcal{C}^{\bot_E}$, i.e., $G_i=E_i$ for all $i=0,1,\ldots,r+2t$, the following theorem can be derived directly by Theorem 3.1:

\vskip 3mm \noindent
  {\bf Theorem 4.3}  \textit{Let $q$ be even. Then all distinct Euclidean self-dual left $D_{2n}$-codes over $\mathbb{F}_q$ are given by Eq. (\ref{eq2}) in Theorem 3.1,
where $C_i$ is a linear code over $K_i$ of length $2$ with generator matrix $G_i$,
$i\leq i\leq r+2t$, and $G_i$ is given by the following three cases}:

\vskip 2mm\par
  (i) $G_0=(1,1)$.

\vskip 2mm\par
  (ii) \textit{Let $1\leq i\leq r$. Then
$G_i=(\varrho_i(x)^{s(q^{\frac{d_i}{2}}-1)},1)$, where $s$ is an arbitrary integer such that $0\leq s\leq q^{\frac{d_i}{2}}$}.

\vskip 2mm\par
  (iii) \textit{Let $i=r+j$ where $1\leq j\leq t$. Then the pairs $(G_i,G_{i+t})$ of matrices
are given by the following two cases}:

\par
  $\diamond$ \textit{$G_i=(1,0)$ and $G_{i+t}=(0,1)$};

\par
  $\diamond$ \textit{$G_i=(g(x),1)$ and $G_{i+t}=(1,g(x^{-1}))$, where
$g(x)=\sum_{l=0}^{d_i-1}a_lx^l$ and $g(x^{-1})=a_0+\sum_{l=1}^{d_i-1}a_lx^{n-l} \ ({\rm mod} \ f_{i+t}(x))$
for any $a_0,a_1,\ldots, a_{d_i-1}\in \mathbb{F}_q$}.

\vskip 3mm\par
 \textit{Therefore, the number of Euclidean self-dual left $D_{2n}$-codes over $\mathbb{F}_q$ is equal to}
$$\mathcal{N}_{{\rm E-SD}}=\prod_{i=1}^{r}(q^{\frac{d_i}{2}}+1)\prod_{j=1}^{t}(q^{d_{r+j}}+1).$$

\vskip 3mm\noindent
  {\bf Remarks} ($\dag$) When $q$ is odd, by \cite[Page 110]{CaoFFA2016}, we know that
there is no Euclidean self-dual left $D_{2n}$-codes over $\mathbb{F}_q$.

  ($\ddag$) When $q$ is even, by Theorems 3.1 and 4.3, we have
$$\frac{\mathcal{N}_{{\rm E-SD}}}{\mathcal{N}}=\frac{1}{3}\prod_{i=1}^{r}\left(1-\frac{2}{q^{\frac{d_i}{2}}+3}\right)
\prod_{j=1}^t\left(1-\frac{2}{q^{d_{r+j}}+3}\right)\approx \frac{1}{3},$$
if all $q^{d_{i}}$ and $q^{d_{r+j}}$ are large enough.

  Hence the probability of a left $D_{2n}$-code over $\mathbb{F}_q$ being an Euclidean self-dual code is approximately equal to $\frac{1}{3}$, when $q$ is even (and large enough).

\vskip 3mm
   Let $\mathcal{N}_{{\rm E-SD}(n,2^m)}$ be the number of Euclidean self-dual left $D_{2n}$-codes over the finite field $\mathbb{F}_{2^m}$. Then for $m\in\{1,2,3,4\}$ and $n\in\{5,7,9,11,13,
15,17\}$, we have the following table:
\begin{center}
\begin{tabular}{l|llll}\hline
 $n$ & $\mathcal{N}_{{\rm E-SD}(n,2)}$  & $\mathcal{N}_{{\rm E-SD}(n,4)}$ &  $\mathcal{N}_{{\rm E-SD}(n,8)}$ & $\mathcal{N}_{{\rm E-SD}(n,16)}$ \\ \hline
$5$ & $5$ & $25$ & 65 & $289^{\star}$\\
$7$ & $9$ & $65$  & $729^{\star}$ & $4097$\\
$9$  & $27$ & $325$ & $6561^{\star}$ & $69649^{\star}$ \\
$11$ & $33$ & $1025$ & $32769$ & $1048577$ \\
$13$  & $65$ & $4225$ & $274625$ & $16785409$\\
$15$ & $255$ & $36125^{\star}$ & $2396745$  & $410338673^{\star}$  \\
$17$ & $289$ & $83521$ & $16785409$ & $6975757441$  \\
\hline
\end{tabular}
\end{center}
\noindent
 This table is from the table of example 6.3 in \cite{CaoFFA2016}, but there are six errors
 (two of them are typos).
Here, we put $^{\star}$ to the six corrected values.

\section{Encoder (generator matrix) of any left $D_{2n}$-code}
\noindent
  In this section, we discuss in detail how to concretely construct each distinct
left $D_{2n}$-code (resp. Euclidean self-dual left $D_{2n}$-code, Euclidean LCD left $D_{2n}$-code
and Euclidean self-orthogonal left $D_{2n}$-code) over $\mathbb{F}_{q}$.

\par
  Vedenev and Deundyak constructed
$\mathbb{F}_q$-basis, generating and check matrices of any dihedral code which are based the representation of
the dihedral codes given in \cite{Vedenev20}. Five pages (Pages 12--16) were used to give their results. However, these results are too complex to easily use to construct specific dihedral codes.

   Here, we provide a direct and simple method for determining the encoder (generator matrix) of any left $D_{2n}$-code over $\mathbb{F}_{q}$. To do this, in the rest of this paper, we identify each polynomial
$$a(x)=a_0+a_1x+\ldots+a_{n-1}x^{n-1}\in\mathcal{A}=\mathbb{F}_{q}[x]/\langle x^{n}-1\rangle$$
with the vector $(a_0,a_1,\ldots,a_{n-1})\in \mathbb{F}_{q}^{n}$. Moreover, for any integer $k$: $1\leq k\leq n-1$, we
set:
$$
[a(x)]_k=\left(\begin{array}{c}a(x)\cr xa(x)\cr \ldots \cr x^{k-1}a(x)\end{array}\right)
=\left(\begin{array}{ccccc}a_0 & a_1 &\ldots & a_{n-2} & a_{n-1}\cr
a_{n-1} & a_0 &\ldots & a_{n-3} & a_{n-2}\cr
\ldots &\ldots &\ldots &\ldots &\ldots \cr
a_{n-k+1} & a_{n-k+2} & \ldots & a_{n-k-1} & a_{n-k}\end{array}\right)
$$
in which the operation is done modulo $x^n-1$. Then $[a(x)]_k\in
{\rm M}_{k\times n}(\mathbb{F}_q)$.

   Now, an explicit generator matrix of each left $D_{2n}$-code over $\mathbb{F}_q$ is given by the
following theorem.

\vskip 3mm\noindent
  {\bf Theorem 5.1} \textit{Let $\mathcal{C}$ be any left $D_{2n}$-code over $\mathbb{F}_q$
with canonical representation $\mathcal{C}=\bigoplus_{i=0}^{r+2t}(\mathcal{A}_i\Box_{\varphi_i}C_i)$, where $C_i$ is a linear code of length $2$ with a generator matrix $G_i$ given
by Theorem 3.1 (resp. Theorem 4.1, Theorem 4.2, or Theorem 4.3). Then}
$${\rm dim}_{\mathbb{F}_q}(\mathcal{C})=\sum_{i=0}^{r+2t}d_i\cdot {\rm dim}_{K_i}(C_i)
=\sum_{i=0}^{r}d_i\cdot {\rm dim}_{K_i}(C_i)+2\sum_{i=r+1}^{r+t}d_i\cdot {\rm dim}_{K_i}(C_i)$$
\textit{and an $\mathbb{F}_q$-generator matrix $G_{\mathcal{C}}$ of $\mathcal{C}$ is given as below}:
$$G_{\mathcal{C}}=\left(\begin{array}{c}G_{\mathcal{A}_0\Box_{\varphi_i}C_0}\cr
G_{\mathcal{A}_1\Box_{\varphi_1}C_1} \cr
\vdots \cr G_{\mathcal{A}_{r+2t}\Box_{\varphi_{r+2t}}C_{r+2t}}\end{array}\right),$$
\textit{where $G_{\mathcal{A}_i\Box_{\varphi_i}C_i}$ is a generator matrix of the concatenated code
$\mathcal{A}_i\Box_{\varphi_i}C_i$, $0\leq i\leq r+2t$,
and the matrix $G_{\mathcal{A}_i\Box_{\varphi_i}C_i}$ is determined
by one of the following three cases}:

\vskip 2mm\par
  1) \textit{Let $G_i=(0,0)$. Then $G_{\mathcal{A}_i\Box_{\varphi_i}C_i}=(0,0,\ldots,0)\in \mathbb{F}_{q}^{2n}$}.

\vskip 2mm\par
  2) \textit{Let $G_i=(g_1(x),g_2(x))$, where
$(g_1(x),g_2(x))\in K_i^2\setminus\{(0,0)\}$. Then}
$$G_{\mathcal{A}_i\Box_{\varphi_i}C_i}=\left([\varepsilon_i(x)g_1(x)]_{d_i},[\varepsilon_i(x)g_2(x)]_{d_i}\right)
\in {\rm M}_{d_i\times 2n}(\mathbb{F}_q).$$

\par
  3) \textit{Let $G_i=I_2$. Then}
$G_{\mathcal{A}_i\Box_{\varphi_i}C_i}=\left(\begin{array}{cc}[\varepsilon_i(x)]_{d_i} & 0 \cr
0 & [\varepsilon_i(x)]_{d_i}\end{array}\right)\in {\rm M}_{2d_i\times 2n}(\mathbb{F}_q).$

\vskip 2mm\textit{Therefore, $\mathcal{C}=\{(u_1,\ldots,u_k)G_{\mathcal{C}}\mid u_1,\ldots,u_k\in \mathbb{F}_q\}$,
where $k={\rm dim}_{\mathbb{F}_q}(\mathcal{C})$}.

\vskip 3mm\noindent
  {\bf Proof.} Let $i$ be an integer: $0\leq i\leq r+2t$, and let $G_i$ be given
by Theorem 3.1, which is the generator matrix
of the outer code $C_i$ in the concatenated code $\mathcal{A}_i\Box_{\varphi_i}C_i$.
By Theorem 3.1, we have one of the following three cases:

\par
  \textsl{Case 1.} Let $G_i=(0,0)$. Then $C_i=\{0\}$ and hence
$\mathcal{A}_i\Box_{\varphi_i}C_i=\{(0,\ldots,0)\}$ $\subseteq \mathbb{F}_{q}^{2n}$. The latter implies  $G_{\mathcal{A}_i\Box_{\varphi_i}C_i}=(0,0,\ldots,0)\in \mathbb{F}_{q}^{2n}$.

\par
  \textsl{Case 2.} Let $G_i=(g_1(x),g_2(x))$, where
$(g_1(x),g_2(x))\in K_i^2\setminus\{(0,0)\}$. Then ${\rm dim}_{K_i}(C_i)=1$
and hence $C_i=\{(\alpha g_1(x),\alpha g_2(x))\mid \alpha\in K_i\}$. From this, by
the definition of concatenated codes in Section 2 and
$K_i=\frac{\mathbb{F}_q[x]}{\langle f_i(x)\rangle}=\{\sum_{j=0}^{d_i-1}b_jx^j\mid b_0,b_1,\ldots,b_{d_i-1}\in\mathbb{F}_q\}$,
we deduce that
\begin{eqnarray*}
 && \mathcal{A}_i\Box_{\varphi_i}C_i \\
 &=& \{(\varepsilon_i(x) \xi(x),\varepsilon_i(x) \eta(x))\mid (\xi(x),\eta(x))\in C_i\} \\
 &=& \{(\sum_{j=0}^{d_i-1}b_jx^j\varepsilon_i(x)g_1(x),\sum_{j=0}^{d_i-1}b_jx^j\varepsilon_i(x)g_1(x))\mid b_j\in \mathbb{F}_q,
 0\leq j\leq d_i-1\}\\
 &=&\{\sum_{j=0}^{d_i-1}b_j(x^j\varepsilon_i(x)g_1(x),x^j\varepsilon_i(x)g_2(x))\mid b_j\in \mathbb{F}_q,
 0\leq j\leq d_i-1\} \\
 &=&\{(b_0,b_1,\ldots,b_{d_i-1})\cdot G_{\mathcal{A}_i\Box_{\varphi_i}C_i}\mid b_j\in \mathbb{F}_q,
 0\leq j\leq d_i-1\},
\end{eqnarray*}
where $G_{\mathcal{A}_i\Box_{\varphi_i}C_i}=\left([\varepsilon_i(x)g_1(x)]_{d_i},[\varepsilon_i(x)g_2(x)]_{d_i}\right)\in
{\rm M}_{d_i\times 2n}(\mathbb{F}_q)$. Hence $G_{\mathcal{A}_i\Box_{\varphi_i}C_i}$ is a generator
matrix of the $\mathbb{F}_q$-linear code $\mathcal{A}_i\Box_{\varphi_i}C_i$.

\par
  \textsl{Case 3.} Let $G_i=I_2$. In this case, we have that
$$\mathcal{A}_i\Box_{\varphi_i}C_i=(\mathcal{A}_i\Box_{\varphi_i}C_i^{(1)})\oplus (\mathcal{A}_i\Box_{\varphi_i}C_i^{(2)}),$$
where $C_i^{(1)}$ and $C_i^{(2)}$ are linear codes over $K_i$ with generator matrices $(1,0)$ and $(0,1)$ respectively.
Using the conclusion of Case 2, $\left([\varepsilon_i(x)]_{d_i},0\right)$ and
$\left(0,[\varepsilon_i(x)]_{d_i}\right)$ are generator matrices of the $\mathbb{F}_q$-linear
codes $\mathcal{A}_i\Box_{\varphi_i}C_i^{(1)}$ and $\mathcal{A}_i\Box_{\varphi_i}C_i^{(2)}$, respectively. Hence
$G_{\mathcal{A}_i\Box_{\varphi_i}C_i}=\left(\begin{array}{cc}[\varepsilon_i(x)]_{d_i} & 0 \cr
0 & [\varepsilon_i(x)]_{d_i}\end{array}\right)$ is a generator matrix of the $\mathbb{F}_q$-linear
codes $\mathcal{A}_i\Box_{\varphi_i}C_i$.

\par
  Finally, by $\mathcal{C}=\bigoplus_{i=0}^{r+2t}(\mathcal{A}_i\Box_{\varphi_i}C_i)$, we conclude that
a generator matrix of the left $D_{2n}$-code $\mathcal{C}$ over $\mathbb{F}_q$ is given by
$G_{\mathcal{C}}=\left(\begin{array}{c}G_{\mathcal{A}_0\Box_{\varphi_i}C_0}\cr
G_{\mathcal{A}_1\Box_{\varphi_1}C_1} \cr
\vdots \cr G_{\mathcal{A}_{r+2t}\Box_{\varphi_{r+2t}}C_{r+2t}}\end{array}\right)$.
\hfill $\Box$

\vskip 3mm\noindent
   {\bf Example 5.2} We consider binary self-dual left $D_{42}$-codes and
binary self-orthogonal left $D_{42}$-codes. In $\mathbb{F}_2[x]$, we have
$$x^{21}-1=f_0(x)f_1(x)f_2(x)f_3(x)f_4(x)f_5(x),$$
where $f_0(x)=x+1$, $f_1(x)=x^2+x+1$, $f_2(x)=x^3+x+1$, $f_3(x)=x^6+x^5+x^4+x^2+1$,
$f_4(x)=x^3+x^2+1$ and $f_5(x)=x^6+x^4+x^2+x+1$ satisfying:
$f_i^\ast(x)=f_i(x)$ for $i=0,1$;
$f_{1+j}^{\ast}(x)=f_{3+j}(x)$ for $i=1,2$. Hence $r=1$, $t=2$, $d_1=2$, $d_2=d_4=3$ and $d_3=d_5=6$.
By Theorems 3.1, 4.3 and 4.2, the number of binary left $D_{42}$-codes
is equal to
$$\mathcal{N}=3\cdot(2^{\frac{2}{2}}+3)(2^3+3)(2^6+3)=11055,$$
the number of binary self-dual left $D_{42}$-codes is equal to
$$\mathcal{N}_{E-SD}=(2^{\frac{2}{2}}+1)(2^3+1)(2^6+1)=1365,$$
and the number of binary self-orthogonal left $D_{42}$-codes is equal to
$$\mathcal{N}_{E-SO}=2\cdot(2^{\frac{2}{2}}+2)(2^3+2)(2^6+2)=5280.$$

\par
  Using the notation of Section 2, we have

\par
  $\varepsilon_0(x)=\sum_{j=0}^{20}x^j$;

\par
  $\varepsilon_1(x)=x+x^2+x^4+x^5+x^7+x^8+x^{10}+x^{11}+x^{13}+x^{14}+x^{16}+x^{17}+x^{19}+x^{20}$;

\par
  $\varepsilon_2(x)=1+x+x^2+x^4+x^7+x^8+x^9+x^{11}+x^{14}+x^{15}+x^{16}+x^{18}$;

\par
  $\varepsilon_3(x)=x^5+x^7+x^{10}+x^{13}+x^{14}+x^{17}+x^{19}+x^{20}$;

\par
  $\varepsilon_4(x)=1+x^3+x^5+x^6+x^7+x^{10}+x^{12}+x^{13}+x^{14}+x^{17}+x^{19}+x^{20}$;

\par
  $\varepsilon_5(x)=x+x^2+x^4+x^7+x^8+x^{11}+x^{14}+x^{16}$.

\par
  Let $\varrho_1(x)=x$ which is a primitive element of the finite
field $K_1=\frac{\mathbb{F}_2[x]}{\langle x^2+x+1\rangle}$ Then $\varrho_1(x)^{2^{\frac{d_1}{2}}-1}=x$
and
$\{\varrho_1(x)^{s(2^{\frac{d_1}{2}}-1)}\mid s=0,1,2=2^{\frac{d_1}{2}}\}=\{1,x,1+x\}$.

\par
  By Theorems 4.2 and 5.1, all distinct $5280$ binary self-orthogonal left $D_{42}$-codes
are generated by the following matrices:
$G=\left(\begin{array}{c} G_0 \cr G_1 \cr \vdots \cr G_5\end{array}\right)$,
where

\par
  $\diamond$ $G_0\in \{(0,0,\ldots,0), ([\varepsilon_0(x)]_1, [\varepsilon_0(x)]_1)\}$.

\par
  $\diamond$ $G_1=\{(0,0,\ldots,0)\}\cup\{([c(x)\varepsilon_1(x)]_2, [\varepsilon_1(x)]_2)\mid c(x)\in {1,x,1+x}\}$.

\par
  $\diamond$ The pairs $(G_2,G_4)$ of matrices are given by the following three cases:
\begin{description}
\item
 $G_2=G_4=(0,0,\ldots,0)$;

\item
  $G_2=([\varepsilon_2(x)]_3, 0)$ and $G_4=(0, [\varepsilon_4(x)]_3)$;

\item
  $G_2=([g_2(x)\varepsilon_2(x)]_3, [\varepsilon_2(x)]_3)$ and $G_4=([\varepsilon_4(x)]_3, [g_2(x^{-1})\varepsilon_4(x)]_3)$, where
\begin{description}
\item{$\checkmark$}
  $g_2(x)=a_0+a_1x+a_2x^2$,

\item{$\checkmark$}
  $g_2(x^{-1})=g_2(x^{20})=a_0+a_2+(a_1+a_2)x+a_1x^2$ (mod $f_4(x)$),
\end{description}
 for any arbitrary $a_0,a_1,a_2\in \mathbb{F}_2$.
\end{description}
\par
  $\diamond$ The pairs $(G_3,G_5)$ of matrices are given by the following three cases:

\begin{description}
\item
  $G_3=G_5=(0,0,\ldots,0)$;

\item
  $G_3=([\varepsilon_3(x)]_6, 0)$ and $G_5=(0, [\varepsilon_5(x)]_6)$;

\item
  $G_3=([g_3(x)\varepsilon_3(x)]_6, [\varepsilon_3(x)]_6)$ and $G_5=([\varepsilon_5(x)]_3, [g_3(x^{-1})\varepsilon_5(x)]_6)$, where
\begin{description}
\item{$\checkmark$}
  $g_3(x)=b_0+b_1x+b_2x^2+b_3x^3+b_4x^4+b_5x^5$,

\item{$\checkmark$} $g_3(x^{-1})=g_3(x^{20})$ (mod $f_5(x)$), i.e.,

 $g_3(x^{-1}) = b_0+b_1+b_3+(b_1+b_2+b_3+b_5)x+(b_2+b_3+b_4)x^2$ $+(b_1+b_2+b_3)x^3+(b_2+b_3+b_5)x^4+(b_1+b_2+b_4)x^5$,
\end{description}
 for any arbitrary $b_j\in \mathbb{F}_2$ and $j=0,1,2,3,4,5$.
\end{description}

\vskip 3mm\noindent
   {\bf Example 5.3} We consider the construction of  LCD left $D_{20}$-codes over $\mathbb{F}_3$.
   In $\mathbb{F}_3[x]$, we have
$$x^{10}-1=(x-1)(x+1)(x^4+x^3+x^2+x+1)(x^4+2x^3+x^2+2x+1).$$
Here $r=3$, $t=0$, $\mathcal{I}_0=\{0,1\}$, $f_0(x)=x-1$, $f_1(x)=x+1$,
$f_2(x)=x^4+x^3+x^2+x+1=f_2^\ast(x)$, $f_3(x)=x^4+2x^3+x^2+2x+1=f_3^\ast(x)$,
$d_0=d_1=1$ and $d_2=d_3=4$. Since $t=0$, by Theorem 4.1, every left $D_{20}$-codes over $\mathbb{F}_3$
must be a LCD code, and
the number of LCD left $D_{20}$-codes over $\mathbb{F}_3$
is equal to
$\mathcal{N}_{E-LCD}=4^2\cdot(3^{\frac{4}{2}}+3)^2=2304.$

\par
  Using the notation of Section 2, we have

\par
  $\varepsilon_0(x)=1+x+x^2+x^3+x^4+x^5+x^6+x^7+x^8+x^9$;

\par
  $\varepsilon_1(x)=1+2x+x^2+2x^3+x^4+2x^5+x^6+2x^7+x^8+2x^9$;

\par
  $\varepsilon_2(x)=1+2x+2x^2+2x^3+2x^4+x^5+2x^6+2x^7+2x^8+2x^9$;

\par
  $\varepsilon_3(x)=1+x+2x^2+x^3+2x^4+2x^5+2x^6+x^7+2x^8+x^9$.

\par
  Let $\varrho_2(x)=1+2x$ which is a primitive element of the finite
field $K_2=\frac{\mathbb{F}_3[x]}{\langle x^4+x^3+x^2+x+1\rangle}$. Then $\varrho_2(x)^{3^{\frac{d_2}{2}}-1}=(1+2x)^8
=1+x+x^2+x^3$.

\par
  Let $\varrho_3(x)=1+x$ which is a primitive element of the finite
field $K_3=\frac{\mathbb{F}_3[x]}{\langle x^4+2x^3+x^2+2x+1\rangle}$. Then $\varrho_3(x)^{3^{\frac{d_3}{2}}-1}=(1+x)^8
=1+2x+x^2+2x^3$.

\par
  Set $\Xi_i=\{(0,0,\ldots,0), \left(\begin{array}{cc}[\varepsilon_i(x)]_{d_i} & 0 \cr
0 & [\varepsilon_i(x)]_{d_i}\end{array}\right)\}$, where $i=0,1,2,3$.

\par
  By Theorems 4.1 and 5.1, all distinct $2304$ LCD left $D_{20}$-codes over $\mathbb{F}_3$
are generated by the following matrices:
$G=\left(\begin{array}{c} G_0 \cr G_1 \cr G_2 \cr G_3\end{array}\right)$,
where

\par
  $\diamond$ $G_i\in \Xi_i\cup\{([\varepsilon_i(x)]_{1}, [\varepsilon_i(x)]_{1}), (-[\varepsilon_i(x)]_{1}, [\varepsilon_i(x)]_{1})\}$, for $i=0,1$.

\vskip 2mm\par
  $\diamond$ $G_i\in \Xi_i\cup\{([\varrho_i(x)^{8s}\varepsilon_i(x)]_{4}, [\varepsilon_i(x)]_{4})\mid s=0,1,2,3,4,5,6,7,8,9\}$,
for $i=2,3$.

\vskip 3mm\noindent
   {\bf Example 5.4} We consider the construction of  LCD left $D_{26}$-codes over $\mathbb{F}_3$. In $\mathbb{F}_3[x]$,
we have
$$x^{13}-1=f_0(x)f_1(x)f_2(x)f_3(x)f_4(x),$$
where $f_0(x)=x-1$, $f_1(x)=x^3+x^2+2$, $f_2(x)=x^3+x^2+x+2$,
$f_3(x)=f_1^\ast(x)=x^3+2x+2$ and $f_4(x)=f_2^\ast(x)=x^3+2x^2+2x+2$.
Hence $r=0$, $t=2$, $d_0=1$ and $d_1=d_2=d_3=d_4=3$. By Theorem 3.1 and 4.1, the number of
left $D_{26}$-codes over $\mathbb{F}_3$ is equal to
$\mathcal{N}=4\cdot(3^{3}+3)^2=3600,$
and the number of LCD left $D_{26}$-codes over $\mathbb{F}_3$
is equal to
$$\mathcal{N}_{E-LCD}=4\cdot(3^{3}+1)^2=3136.$$

\par
  Using the notation of Section 2, we have

\par
  $\varepsilon_0(x)=1+x+x^2+x^3+x^4+x^5+x^6+x^7+x^8+x^9+x^{10}+x^{11}+x^{12}$;

\par
  $\varepsilon_1(x)=2x^2+2x^4+2x^5+2x^6+x^7+x^8+2x^{10}+x^{11}+2x^{12}$;

\par
  $\varepsilon_2(x)=x+x^3+2x^4+2x^7+2x^8+x^9+2x^{10}+2x^{11}+2x^{12}$;

\par
  $\varepsilon_3(x)=2x+x^2+2x^3+x^5+x^6+2x^7+2x^8+2x^9+2x^{11}$;

\par
  $\varepsilon_4(x)=2x+2x^2+2x^3+x^4+2x^5+2x^6+2x^9+x^{10}+x^{12}$.

\par
  By Theorems 4.1 and 5.1, all distinct $3136$ LCD left $D_{26}$-codes over $\mathbb{F}_3$
are generated by the following matrices:
$G=\left(\begin{array}{c} G_0 \cr G_1 \cr G_2 \cr G_3 \cr G_4\end{array}\right)$,
where

\par
  $\diamond$ $G_0\in \{(0,\ldots,0), \left(\begin{array}{cc}[\varepsilon_0(x)]_{1} & 0 \cr
0 & [\varepsilon_0(x)]_{1}\end{array}\right), ([\varepsilon_0(x)]_{1}, [\varepsilon_0(x)]_{1}), (-[\varepsilon_0(x)]_{1}$, $[\varepsilon_0(x)]_{1})\}$.

\par
  $\diamond$ The pairs $(G_1,G_3)$ of matrices are given by the following three cases:
\begin{description}
\item
 $G_1=G_3=(0,0,\ldots,0)$;

\item
  $G_1=\left(\begin{array}{cc}[\varepsilon_1(x)]_{3} & 0 \cr
0 & [\varepsilon_1(x)]_{3}\end{array}\right)$ and $G_3=\left(\begin{array}{cc}[\varepsilon_3(x)]_{3} & 0 \cr
0 & [\varepsilon_3(x)]_{3}\end{array}\right)$;

\item
  $G_1=([g_1(x)\varepsilon_1(x)]_3, [\varepsilon_1(x)]_3)$ and $G_3=([\varepsilon_3(x)]_3, [g_1(x^{-1})\varepsilon_3(x)]_3)$, where
\begin{description}
\item{$\checkmark$}
  $g_1(x)=a_0+a_1x+a_2x^2$,

\item{$\checkmark$}
  $g_1(x^{-1})=g_1(x^{12})=a_0+2a_1+a_2+a_2x+(a_1+2a_2)x^2$ (mod $f_3(x)$),
\end{description}

  for any arbitrary $a_0,a_1,a_2\in \mathbb{F}_3$ satisfying $(a_0,a_1,a_2)\neq (0,0,0)$.
\end{description}
\par
  $\diamond$ The pairs $(G_2,G_4)$ of matrices are given by the following three cases:
\begin{description}
\item
 $G_2=G_4=(0,0,\ldots,0)$;

\item
  $G_2=\left(\begin{array}{cc}[\varepsilon_2(x)]_{3} & 0 \cr
0 & [\varepsilon_2(x)]_{3}\end{array}\right)$ and $G_4=\left(\begin{array}{cc}[\varepsilon_4(x)]_{3} & 0 \cr
0 & [\varepsilon_4(x)]_{3}\end{array}\right)$;

\item
  $G_2=([g_2(x)\varepsilon_2(x)]_3, [\varepsilon_2(x)]_3)$ and $G_4=([\varepsilon_4(x)]_3, [g_2(x^{-1})\varepsilon_4(x)]_3)$, where
\begin{description}
\item{$\checkmark$}
  $g_2(x)=b_0+b_1x+b_2x^2$,

\item{$\checkmark$}
  $g_2(x^{-1})=g_2(x^{12})=b_0+2b_1+2(b_1+b_2)x+(b_1+2b_2)x^2$ (mod $f_4(x)$),
\end{description}
  for any arbitrary $b_0,b_1,b_2\in \mathbb{F}_3$ satisfying $(b_0,b_1,b_2)\neq (0,0,0)$.
\end{description}

\vskip 3mm \noindent
   {\bf Example 5.5} For integers $m\in\{1,2\}$ and
positive integers $n$ satisfying ${\rm gcd}(3,n)=1$ and $4\leq n\leq 20$, using Theorem 4.1, we list the number $\mathcal{N}(n,3^m)$ of
left $D_{2n}$-codes and the number $\mathcal{N}_{E-LCD}(n,3^m)$ of
Euclidean LCD left $D_{2n}$-codes over $\mathbb{F}_{3^m}$ respectively, as a table below:
\begin{center}
\begin{tabular}{l|ll|ll}\hline
 $n$ & $\mathcal{N}(n,3)$  & $\mathcal{N}_{E-LCD}(n,3)$ &  $\mathcal{N}(n,3^2)$ & $\mathcal{N}_{E-LCD}(n,3^2)$ \\ \hline
$4$ & $96$ & $96$ & 192 & 160\\
$5$ & $48$ & $48$ & 576 & 576 \\
$7$ & $120$ & $120$  & $2928$ & $2920$\\
$8$  & $1152$ & $960$ & $27648$ & $16000$ \\
$10$ & $2304$ & $2304$ & $331776$ & $331776$ \\
$11$ & $984$ & $976$ & $236208$ & $236200$ \\
$13$ & $3600$ & $3136$ & $2143296$ & $2131600$\\
$14$ & $14400$ & $14400$ & $8573184$ & $8526400$  \\
$16$ & $96768$ & $78720$ & $195084288$ & $107584000$  \\
$17$ & $26256$ & $26256$ & $172344384$ & $172344384$  \\
$19$ & $78744$ & $78744$ & $1549681968$ & $1549681960$  \\
$20$ & $1161216$ & $1133568$ & $28092137472$ & $22308618240$  \\
\hline
\end{tabular}
\end{center}

\section{Conclusions and further research}
\noindent
We studied the construction and enumeration of left dihedral codes satisfying certain Euclidean duality properties:
LCD left dihedral codes, self-orthogonal left dihedral codes and self-dual left dihedral codes. Specifically, we determined the Euclidean hull of every
left $D_{2n}$-code over $\mathbb{F}_{q}$, where ${\rm gcd}(n,q)=1$, and provided an explicit representation and a precise enumeration for Euclidean LCD, self-orthogonal and self-dual left $D_{2n}$-codes
over $\mathbb{F}_{q}$ respectively. Moreover, we give a direct and simple method for determining the encoder (generator matrix) of any left $D_{2n}$-code over $\mathbb{F}_{q}$.

\par
   Future topics of interest include to determine the Hermitian duality of left dihedral codes and give an explicit
representation and a prices enumeration for left dihedral codes satisfying certain Hermitian duality properties.

\section*{Acknowledgments}

\noindent
This research is
supported in part by the National Natural Science Foundation of
China (Grant Nos. 12071264, 11801324, 11671235), the Shandong Provincial Natural Science Foundation,
China (Grant No. ZR2018BA007) and the Scientific Research Fund of Hubei Provincial Key Laboratory of Applied Mathematics (Hubei University) (Grant Nos. HBAM201906).



\end{document}